\documentclass[twocolumn,aps,eprint,showpacs,superscriptaddress,amsmath,amssymb]{revtex4-1}
\usepackage{mathrsfs}
\usepackage{graphicx}
\usepackage{bm}
\usepackage{float}
\usepackage{physics}
\usepackage{siunitx}
\usepackage{xcolor}
\usepackage{subfigure}
\usepackage{appendix}
\usepackage[colorlinks,linkcolor=blue,anchorcolor=blue,urlcolor=blue,citecolor=blue]{hyperref}
\usepackage{multirow}
\usepackage{multibib}

\begin{document}
\title{ Parallel-Electromagnetically-Induced-Transparency Near Ground-State Cooling  of a  Trapped-ion Crystal}

\author{Jie Zhang }
 \thanks{These authors contributed equally to this work.}

\author{Man-Chao Zhang}
 \thanks{These authors contributed equally to this work.}

\author{Yi Xie}

\author{Chun-Wang Wu}

\author{Bao-Quan Ou}

\author{Ting Chen}
\affiliation{Department of Physics, College of Liberal Arts and Sciences, National	University of Defense Technology, Changsha 410073, Hunan, China.}
\affiliation{Interdisciplinary Center for Quantum Information, National University 	of Defense Technology, Changsha 410073, Hunan, China.}
\affiliation{Hunan Key Laboratory of Quantum Information Mechanism and Technology, National University	of Defense Technology, Changsha 410073, Hunan, China.}
\author{Wan-Su Bao }
\affiliation{Henan Key Laboratory of Quantum Information and Cryptography, Zhengzhou	Information Science and Technology Institute, Zhengzhou 450001, China}
\author{Paul Haljan}
\affiliation{Department of physics, Simon Fraser University, Burnaby, BC, Canada, V5A 1S6}

\author{Wei Wu}
\affiliation{Department of Physics, College of Liberal Arts and Sciences, National
University of Defense Technology, Changsha 410073, Hunan, China.}
\affiliation{Interdisciplinary Center for Quantum Information, National University
of Defense Technology, Changsha 410073, Hunan, China.}
\affiliation{Hunan Key Laboratory of Quantum Information Mechanism and Technology, National University	of Defense Technology, Changsha 410073, Hunan, China.}
\author{Shuo Zhang }
\email{Corresponding author: shuoshuo19851115@163.com}

\affiliation{Henan Key Laboratory of Quantum Information and Cryptography, Zhengzhou
Information Science and Technology Institute, Zhengzhou 450001, China}
\author{Ping-Xing Chen}
\email{Corresponding author: pxchen@nudt.edu.cn}

\affiliation{Department of Physics, College of Liberal Arts and Sciences, National
University of Defense Technology, Changsha 410073, Hunan, China.}
\affiliation{Interdisciplinary Center for Quantum Information, National University
of Defense Technology, Changsha 410073, Hunan, China.}
\affiliation{Hunan Key Laboratory of Quantum Information Mechanism and Technology, National University	of Defense Technology, Changsha 410073, Hunan, China.}

\date{\today}

\begin{abstract}
We theoretically propose and experimentally demonstrate a parallel-electromagnetically-induced-transparency  (parallel-EIT) cooling  technique    for  ion crystals in the Paul trap. It has  less stringent requirements on the cooling resonance condition than the standard electromagnetically-induced-transparency (EIT) cooling, thus allowing, in principle, to simultaneously cool the motional mode spectrum with an arbitrary range. A proof-of-principle validation for this cooling scheme is experimentally demonstrated with up to 4 trapped $^{40}$Ca$^{+}$ ions. We observe simultaneous near-ground-state cooling for all motional modes with best average phonon number about 0.2. By tuning the trap frequency in a large range to imitate a broadband motional mode spectrum, we can still reach almost the same cooling limit for all the modes while standard EIT cooling shows limited cooling range. Our method has a simple experimental configuration, requiring only appropriate modulation of the probe beam of standard EIT cooling, and can be applied to various types of ions (e.g., $^{171}$Yb$^+$, $^{40}$Ca$^+$). This cooling scheme provides a powerful tool for  the initialization of the trapped-ion quantum computers and simulators.



\end{abstract}
\maketitle


Cooling the motional degree of freedom to the ground state
is an essential step for the initialization of trapped-ion based  quantum computers and simulators \cite{PhysRevLett.92.207901,Blatt2012,PhysRevB.89.214305,RevModPhys.85.1103,NP1206,PhysRevLett.74.4091,PhysRevA.100.062111}, where the ground-state cooled motional mode  acts as the bus for establishing  the quantum entanglement between qubits.  In particular,  high fidelity quantum operations \cite{PhysRevLett.117.060504,PhysRevLett.117.140501,wineland1998twoions,PhysRevLett.113.220501} of ion qubits require cooling  the motional modes to the  Lamb-Dicke (LD) regime, in which the vibrational amplitude of each ion is much less than the optical wavelength. With increasing size of trapped-ion crystals,  cooling a large number of  motional modes in one process to near ground state is required to reduce the heating effects from the environment \cite{RevModPhys.87.1419,PhysRevA.62.053807}  and the mutual mode coupling \cite{SaraPHDthesis, RevModPhys.87.1419}.  Particularly the fidelity of quantum operations using transverse mode of a long ion string benifits from the sufficiently cooled axial mode \cite{cetina2020quantum}.  Therefore it is  necessary to exploit  an efficient cooling with high bandwidth to bring  all  motional modes  down to near ground state. 

A typical laser cooling of the trapped ions begins with Doppler cooling
\cite{PhysRevA.20.1521,RevModPhys.58.699,Dalibards, PhysRevLett.127.143201}, followed
by a ground-state cooling technique, including Sisyphus cooling \cite{joshi2020polarization,20173D},
resolved sideband cooling \cite{PhysRevLett.62.403,PhysRevLett.83.4713,PhysRevLett.75.4011},
and electromagnetically-induced-transparency (EIT) cooling \cite{PhysRevLett.85.4458,PhysRevLett.85.5547,PhysRevA.93.053401,qiao2020doubleeit,PhysRevLett.122.053603,PhysRevA.98.023424,PhysRevLett.125.053001,PhysRevLett.110.153002, WebsterPHDthesis,PhysRevA.89.033404}.  While Sisyphus cooling is  efficient  for simultaneous cooling of all motional modes, its cooling limit is  not low enough for ground state preparation \cite{joshi2020polarization,20173D} and it suffers from low cooling rate  for hyperfine qubits \cite{20173D}.  Resolved sideband cooling (RSC)
is a standard ground-state cooling method on a narrow optical transition.
The cooling process takes place for each mode sequentially, hence
the cooling time scales up with numbers of ions. A further improvement
using parallel RSC shows a quadratic speedup for the conventional RSC cooling rate
and a considerable extension for the cooling bandwidth \cite{IonQ2018PRA}. However, it requires simultaneous
single ion addressing technique, which is  very
challenging for large-size ion crystals.

EIT cooling is another widely used ground-state cooling technique.
It employs a two-photon interference in a $\Lambda$ configuration
ion to tailor the absorption profile, which povides a faster method
with a larger cooling bandwidth than RSC \cite{zhang2021fast,PhysRevA.104.043106}, such that several modes
within a certain range can be simultaneously cooled. The state-of-the-art
experiment showed that the EIT cooling bandwidth can be extended up to more than
$\SI{3}{\mega\hertz}$ by reducing the detunings of cooling beams \cite{PhysRevLett.125.053001}.
However, the  cooling bandwidth  is still limited 
for cooling a large-size ion crystals.
To further extend the cooling bandwidth, one can  use more energy levels than the $\Lambda$ structure (e.g., double EIT (ref.\cite{qiao2020doubleeit,PhysRevA.98.023424}),  but this  scheme  depends on the atomic structure and may still not be able to cover all the mode spectrum.


\begin{figure}
\centering
\includegraphics[width=0.48\textwidth]{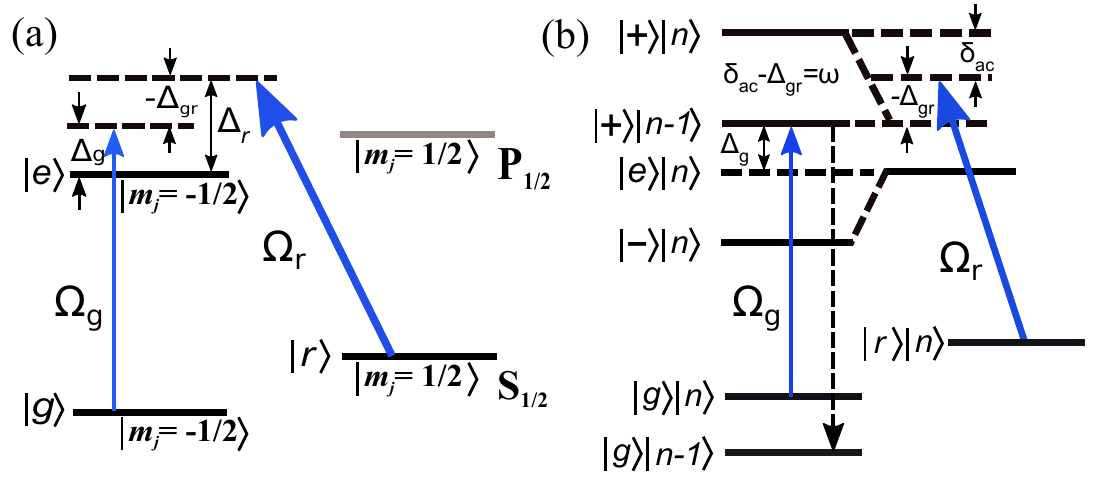}
\caption{(a) The relevant energy levels of  $^{40}$Ca$^{+}$ ion  for implementing the parallel-EIT cooling and EIT cooling. The ion has a dissipative excited state $\ket{e}$  and two ground states $\ket{g}$  and $\ket{r}$ , transitions $\ket{e} \leftrightarrow \ket{g}$  and $\ket{e} \leftrightarrow \ket{r}$  are driven by a probe laser and a driving laser, respectively. When $\Delta_{gr} =0$, the setup is for standard EIT cooling.
(b) The cooling mechanism of parallel-EIT cooling. Under the optimal cooling condition that $\Delta_{gr}-\delta_{ac}=-\omega$, the red sideband transition $\ket{g,n} \leftrightarrow \ket{+, n-1} $ is resonant. }	
\label{fig:figure1}
\end{figure}

In this work, we theoretically propose and experimentally demonstrate a new EIT-like cooling method with arbitrary cooling bandwidth for trapped ions, which we refer as  parallel-EIT cooling method. The level structure and cooling mechanism is similar to standard EIT cooling. However, different from EIT cooling, the cooling resonance can be adjusted by not only the ac Stark shift, but also the detuning difference between the driving beam and probe beam.  This allows us to apply a range of probe laser beams with different detunings for multimode cooling. In this way, all the
motional modes can be simultaneously cooled to ground state, regardless
of their mode separations. Compared with double-EIT cooling \cite{qiao2020doubleeit,PhysRevA.98.023424}, our scheme provides a broader cooling bandwidth and does not require additional enegy levels.  We experimentally demonstrate
the scheme by cooling all  motional modes of 1, 2 and 4 trapped $^{40}$Ca$^{+}$ ions
to near ground state. We prove that our method has  much  wider cooling bandwidth than the EIT cooling by tuning the single mode frequency in a large range to simulate a broadband motional spectrum. Moreover, our method shows no selectivity to the ion species since it  only requires modulations of the probe beam  compared to the EIT cooling.

The relevant levels in parallel-EIT scheme is  same as those
in EIT cooling, as it is shown in Fig. \ref{fig:figure1}(a). An
ion with $\Lambda$ configuration confined in an ion trap has oscillation
frequency $\omega$. The probe laser and driving laser couple to transitions
$\ket{g}\longleftrightarrow\ket{e}$ and $\ket{r}\longleftrightarrow\ket{e}$
with blue detunings $\Delta_{g}$ and $\Delta_{r}$, Rabi frequencies
$\Omega_{g}$ and $\Omega_{r}$, respectively. When $\Omega_{g}\ll\Omega_{r}$,
a narrow-line dressed state $\left|+\right\rangle \approx\frac{\Omega_{r}\left|r\right\rangle +2\delta_{ac}\left|e\right\rangle }{\sqrt{\Omega_{r}^{2}+4\delta_{ac}^{2}}}$
is generated by the strong coupling between $\left|r\right\rangle $
and $\left|e\right\rangle $, with $\delta_{ac}=\frac{1}{2}\left(\sqrt{\Omega_{r}^{2}+\Delta_{r}^{2}}-\left|\Delta_{r}\right|\right)\approx\Omega_{r}^{2}/\left(4\left|\Delta_{r}\right|\right)$
being the ac Stark shift created by driving laser. The effective coupling
$\left|g\right\rangle \longleftrightarrow\left|+\right\rangle $ creates
a Fano-like absorption profile around $\Delta_{r}$, where the absorption
null and the narrow peak are at $\Delta_{r}$ and $\Delta_{r}+\delta_{ac}$,
respectively.

The standard EIT cooling is inspired by the concept
\textquotedblleft coherent population trapping\textquotedblright ,
which is achieved by setting $\Delta_{g}=\Delta_{r}$ and the atomic
excitation vanishes. Meanwhile, by tuning $\delta_{ac}=\omega$, such
that the red sideband aborsption $\left|g\right\rangle \left|n\right\rangle \longrightarrow\left|+\right\rangle \left|n-1\right\rangle $
falls at the peak of the profile, while the off-resonant blue sideband heating is suppressed. As a result, the ion is cooled to
the ground state. However, note that it is not quite necessary to completely eliminate the off-resonant carrier excitation $\left|g\right\rangle \left|n\right\rangle \longrightarrow\left|+\right\rangle \left|n\right\rangle$, for its transition amplitude is negligible compared to that of red sideband transition.  Hence we
drop the constraint of EIT condition that $\Delta_{g}=\Delta_{r}$,
and obtain a more generalized the cooling condition
\begin{equation}
	\Delta_{gr}-\delta_{ac}=-\omega,\label{eqn1}
\end{equation}
with detuning difference $\Delta_{gr}=\Delta_{g}-\Delta_{r}$, where the red sideband transition is resonant as shown in Fig. \ref{fig:figure1}(b). The analytical calculation for parallel-EIT cooling with single probe beam shows that, for an effective Lamb-Dicke (LD)
parameter $\eta$ and natural linewidth $\gamma$ of state $\left|e\right\rangle $,
the optimal cooling rate is $W=\eta^{2}\frac{\Omega_{g}^{2}}{\gamma}$, which is the same as that of standard EIT cooling. Meanwhile, the final mean phonon occupation of parallel EIT $n_{\textrm{st}}= Rn_{\textrm{EIT}}$, where $ R=\left[1+4\alpha\left(1-\frac{\delta_{ac}}{\omega}\right)^{2}\right]$ is the ratio of $n_{\textrm{st}}$ to EIT cooling $n_{\textrm{EIT}} =\frac{\gamma^{2}}{16\Delta_{g}^{2}} $ under the same detuning $\Delta_g$ (see supplement).
\begin{figure}
	\centering
	\includegraphics[width=0.48\textwidth]{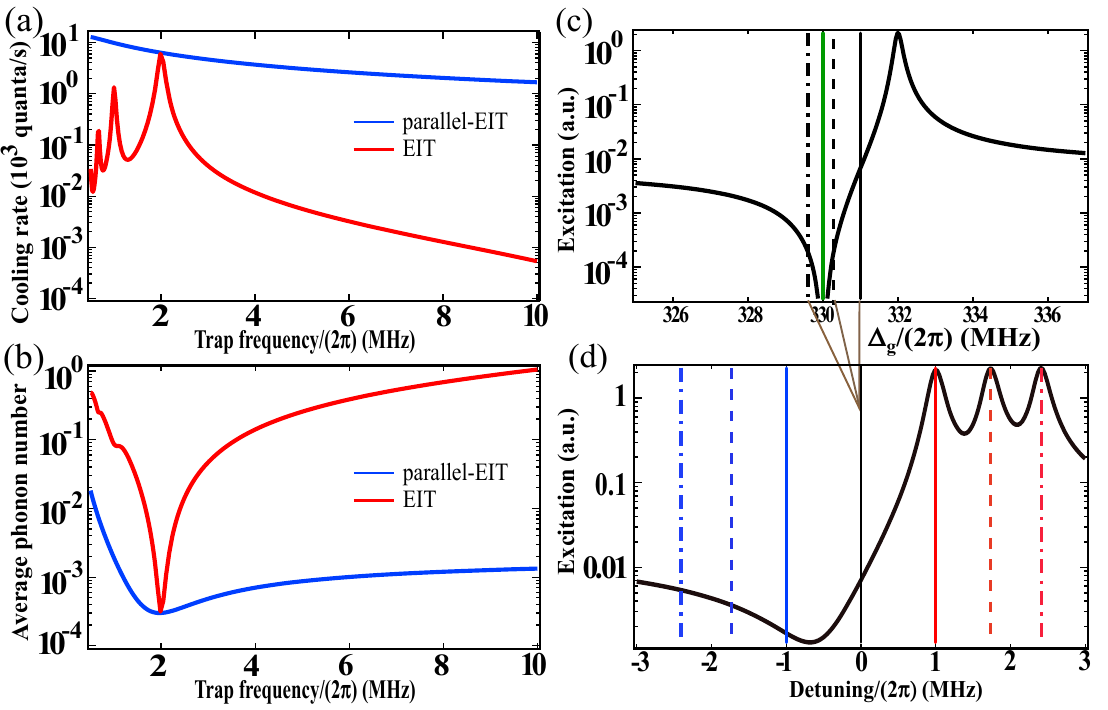}
	\caption{
		(a) and (b) are the simulated final mean phonon numbers and cooling rate for cooling  a single mode  $\omega$ with parallel-EIT cooling and EIT cooling (calculation see supplement). The varied mode frequency simulates the mode spectrum for multiple ions.  The simulation parameters are  $\Delta_r/(2\pi) = 330$ MHz, $\Omega_{g}/(2\pi)= 3$ MHz, and the effective LD parameter $\eta = 0.29/\sqrt{\omega/(2\pi)}$. The ac Stark shift is fixed at $\delta_{ac}/(2\pi) = 2$ MHz. The  probe beam has a constant detuning $\Delta_g = \Delta_r$ for the  EIT cooling, while it is adjusted according to the conditon (\ref{eqn1}) for parallel-EIT cooling.  
		(c) The Fano-like profile around EIT resonance. To simultaneous cool  3 separated modes at $\omega_{1}/(2\pi)=1.0 $ MHz, $\omega_{2}=\sqrt{3}$ MHz and $\omega_{3} = \sqrt{29/5}$ MHz, one can apply 3 probe beams. Each beam is tuned to address  one individual mode. The black vertical lines mark the probe laser detunings for the parallel-EIT cooling and the green vertical line indicates the  detuning for the EIT cooling. 
		 We further align the absorption spectrums of the 3 beams at the carrier transition in (d). The 3 blue (red) vertical lines mark the positions of the blue (red) sidebands.
	}	
	\label{fig:figure2}
\end{figure}

Compared with EIT cooling, parallel-EIT cooling scheme provides a more
flexible cooling resonance. In Fig. \ref{fig:figure2} (a) and (b), we numerically
compare the steady phonon occupations and cooling rates of parallel-EIT cooling and
standard EIT cooling for fixed Fano-like profile. In EIT cooling,
detuning $\Delta_{g}$ has to be tuned equal to $\Delta_{r}$, hence
only the motional mode frequencies around $\delta_{ac}$ can be efficiently
cooled. On the other hand, in parallel-EIT cooling, by tuning $\Delta_{gr}$ to satisfy cooling condition
(\ref{eqn1}), the position of carrier transition can be adjusted, hence
one can optimally cool the motional mode at arbitary frequency. The results also show that parallel-EIT cooling has a robust cooling over a large
range of mode frequencies, which is especially useful for multimode
cooling. Since each mode of an ion string can be cooled independently from the others in LD regime \cite{Morigi_2003}, cooling of multiple motional modes can be realized by applying  a series of  probe beams and one driving beam to cover the whole mode spectrum. The frequency of each probe beam is tuned to meet the optimal condition (\ref{eqn1}) to address one single mode or a range of nearby modes. In this way, all the modes can be optimally cooled.   Fig. \ref{fig:figure2}(c)  shows an example of cooling the axial collective motions of a 3-ion chain using parallel EIT cooling. One tune the frequencies of 3 probe beams such that the red sideband absorption of each mode is around the peak of the profile. To further highlight the cooling bandwidth, we align the positions of carrier absorptions, as shown in Fig.  \ref{fig:figure2}(d). There are three maxima, corresponding to red sideband resonance of the 3 modes.


We experimentally investigate parallel-EIT  cooling method on the dipole transition $S_{1/2}\leftrightarrow P_{1/2}$
of the $^{40}$Ca$^{+}$ ions at the wavelength of 397 nm and a natural
linewidth of $\Gamma=2\pi\times20.7$ MHz. The ions are confined in
a blade-shaped Paul trap with trap frequencies $2\pi\times(0.6-1.4)$
MHz along the axial direction and $ 2\pi\times (1.8-4.02) $  MHz for
the transverse direction depending on the experiments. The relevant
energy levels are formed by introducing a constant magnetic field
$B_{0}=5.49$ G, which is at $\ang{45}$ with respect to the trap axis
 and results in a Zeeman shift of $\pm\Delta_{B}/2\pi=15.36$
MHz between the states $\ket{g}$ and $\ket{r}$.
The $\Lambda$ system for one cooling resonance can be formed by using one of the probe beams and the driving beam, which are tuned  near the two-photon  resonance as depicted in Fig. \ref{fig:figure1}(a). The strong $\sigma^{-}$ light couples the transition $\left|r\right\rangle \longleftrightarrow\left|e\right\rangle $
with blue detuned  $ \Delta_{r} = 2\pi\times 330$ MHz and the weak $\pi$ polarized
light couples the $\left|g\right\rangle \longleftrightarrow\left|e\right\rangle $
with a detuning difference $\Delta_{gr}$ to the $\sigma^{-}$ light.
Multiple cooling resonances can be realized by adding more probe beams, which can be created by modulating probe beam with  AOM  driven by  multiple frequencies (see supplement). In this setup, the wavevector difference $\Delta\vec{k}$ between
the driving and probe beams has components along all motional directions,
therefore our setup can be used for cooling of all motional modes at the same time.

\begin{figure}
\centering %
\begin{minipage}[c]{0.92\linewidth}%
	\centering \includegraphics[width=1\textwidth]{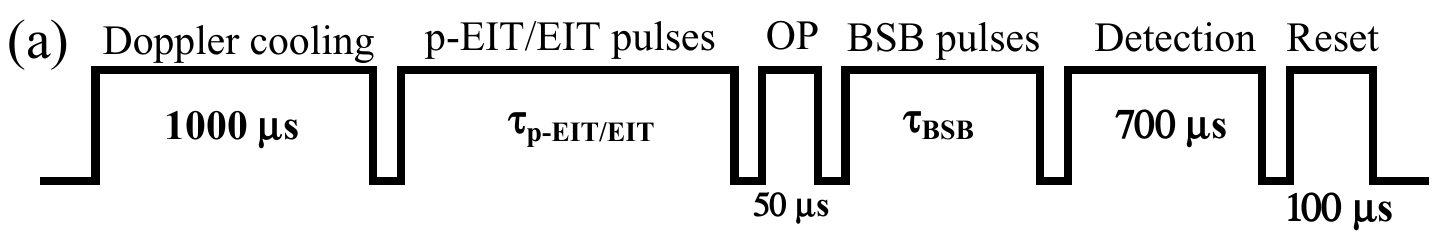} %
\end{minipage}%

\centering %
\begin{minipage}[c]{0.95\linewidth}%
 \centering \includegraphics[width=1\textwidth]{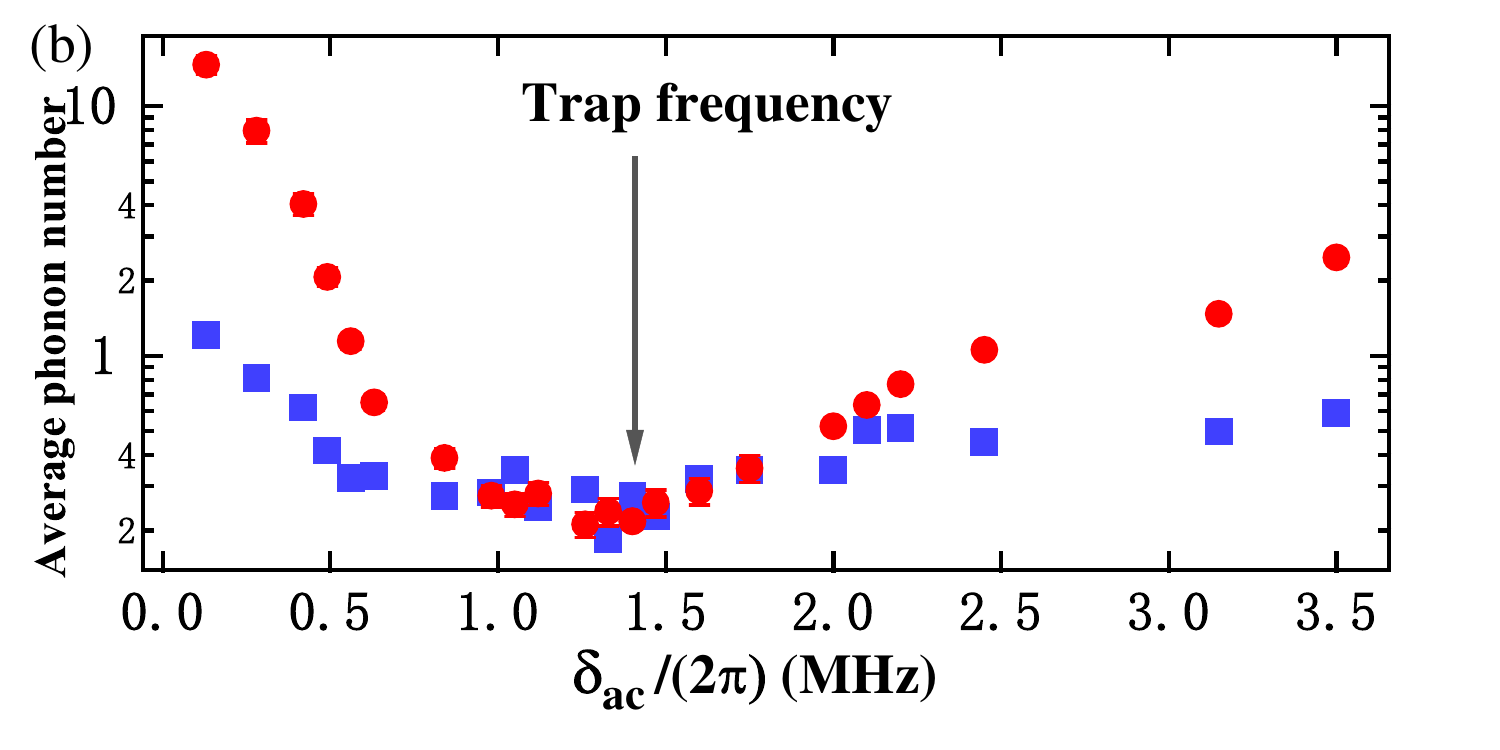} %
\end{minipage}%

\begin{minipage}[c]{0.94\linewidth}%
 \centering \includegraphics[width=1\textwidth]{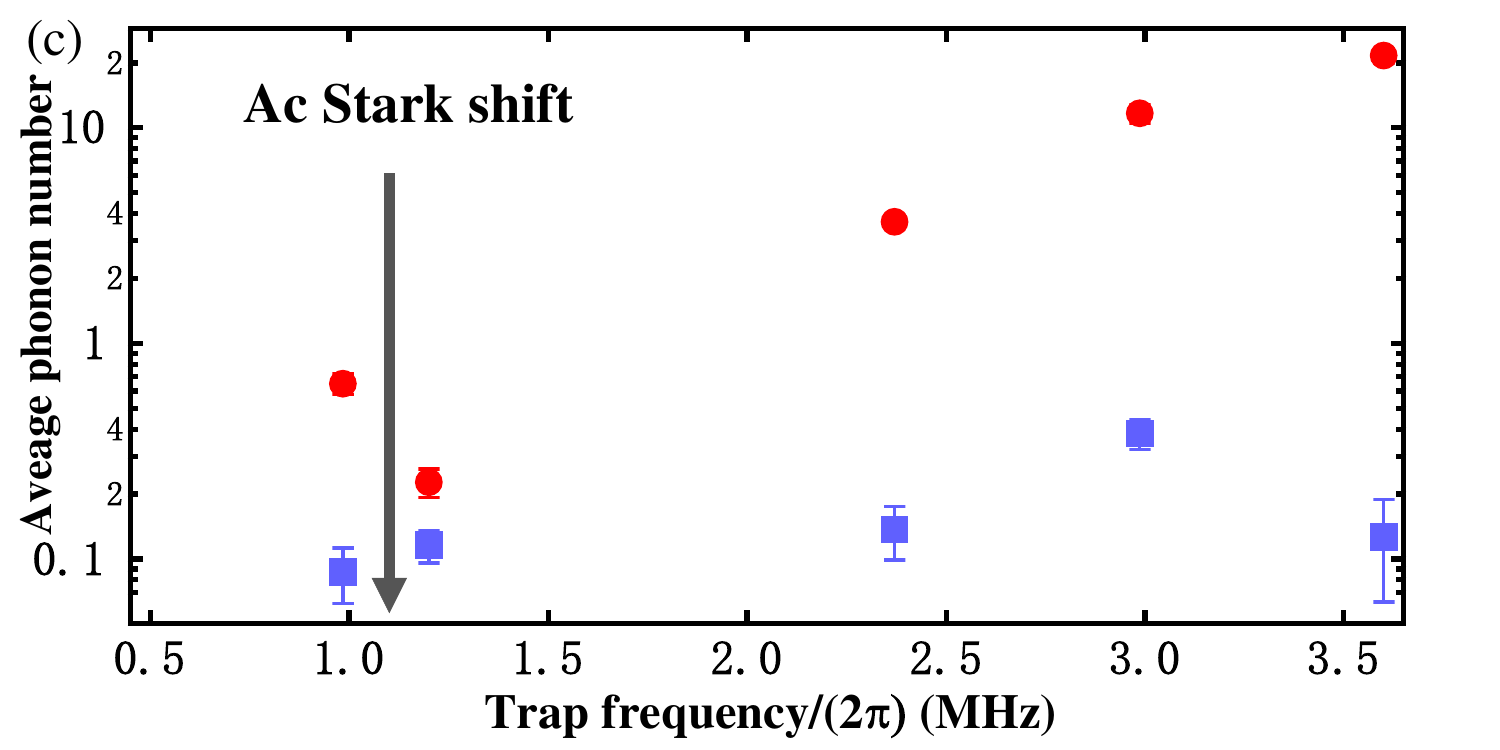} %
\end{minipage}

\caption{(a) Experimental sequence to compare the EIT cooling and parallel-EIT cooling method by using a  single trapped ion. (b) Cooling axial motional mode with frequency  fixed  at $\omega_{z}/2\pi=1.4$
MHz. The blue square and red dots show the cooling results for the
parallel-EIT cooling and EIT cooling respectively.  (c) The cooling results for a fixed
ac Stark shift at $\delta_{ac}/2\pi= \SI{1.2}{\mega\hertz}$ with variable motional frequencies. The parallel-EIT cooling method shows almost same cooling limit for the
varying trap frequencies with best average phonon number about 0.1,
however the EIT cooling limit goes higher as the the trap frequency
deviates from the ac Stark shift.}
\label{fig:figure3}
\end{figure}


\begin{figure}
\centering %
\noindent\begin{minipage}[c]{1\linewidth}%
\centering \includegraphics[width=0.95\textwidth]{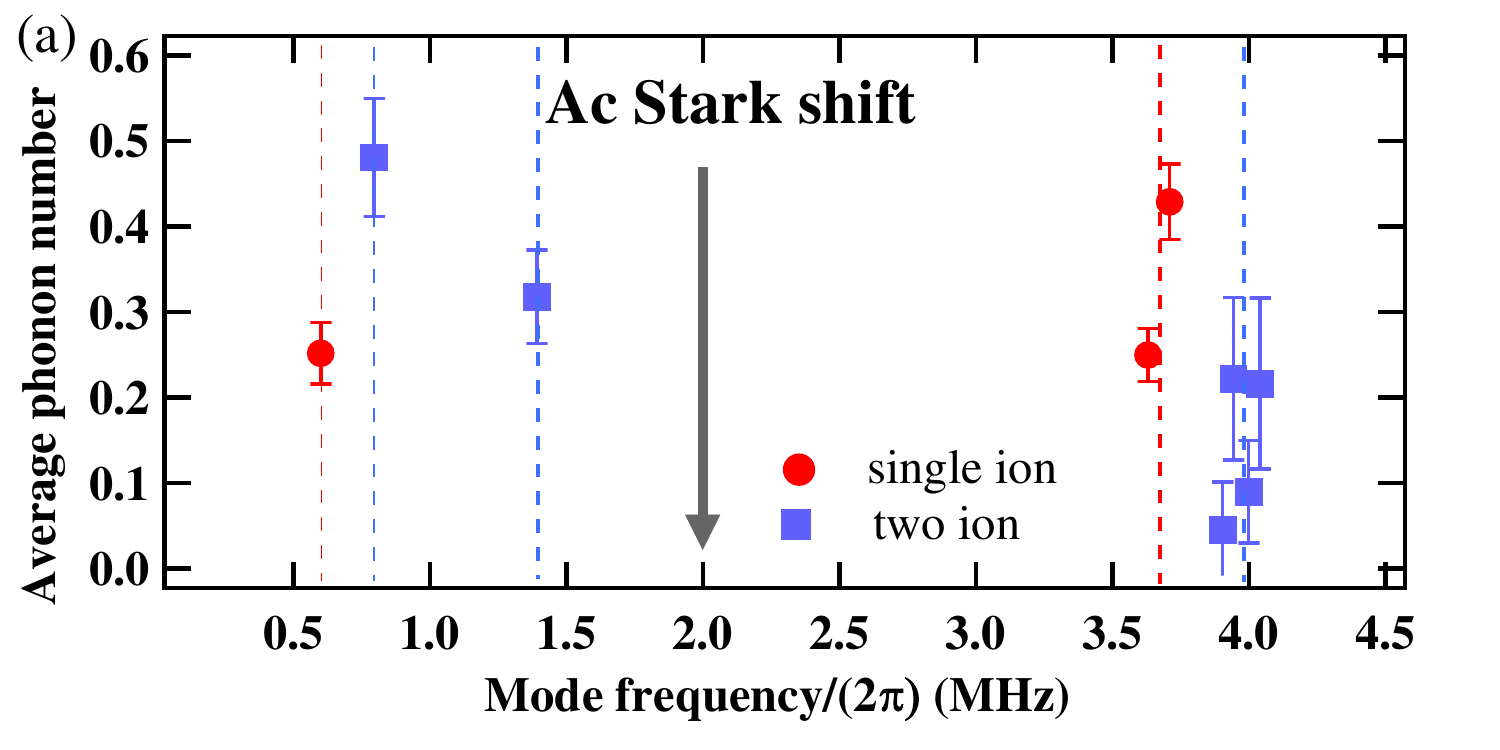}
\includegraphics[width=0.95\textwidth]{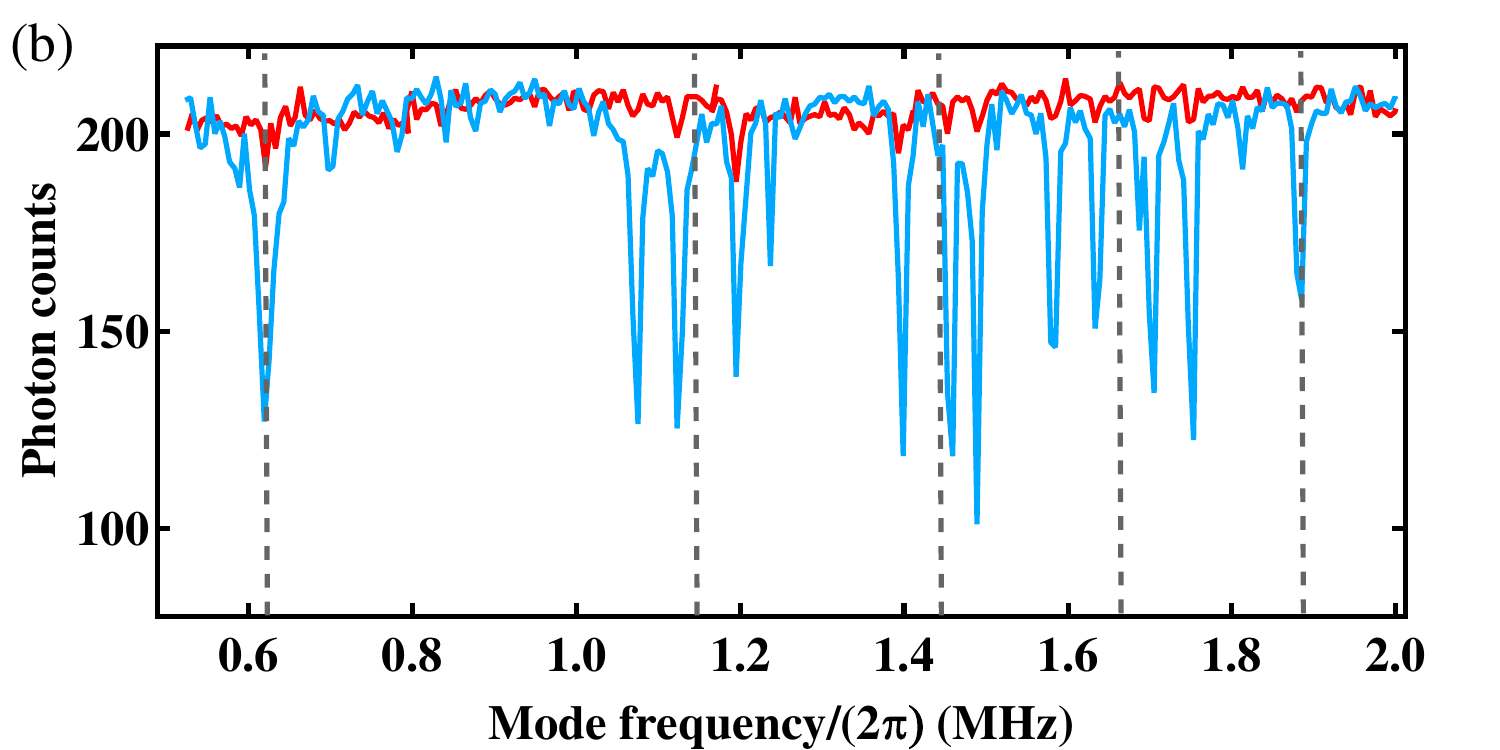}
 \centering \includegraphics[width=0.95\textwidth]{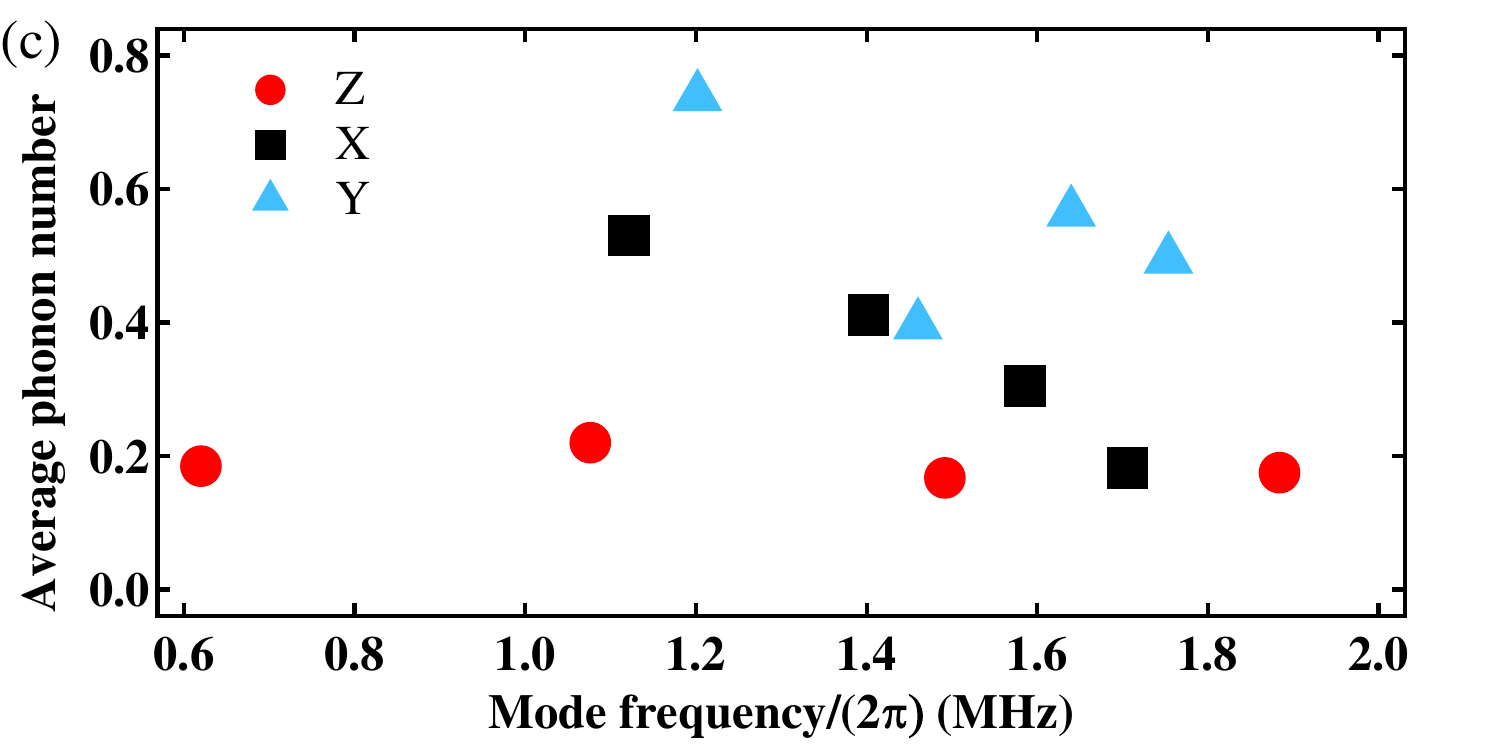} %
\end{minipage}\caption{(a) Mean phonon numbers for all motional modes of one and two $^{40}$Ca$^{+}$ ions after applying a cooling pulse of $\SI{1.5}{\milli\second}$. The vertical arrow shows the  ac Stark shift at around $ 2\pi \times \SI{ 2}{\mega\hertz}$. The red dots and blue squares show the average phonon numbers for the one- and two-ion cases respectively. The vertical dash lines mark  the cooling resonances.
(b) Blue and red sideband spectrum  after applying the cooling pulses for a 4-ion chain.  The horizontal
axis indicates the absolute detuning  between the 729 nm laser  and
the carrier transition $S_{1/2}(m_{j}=-1/2)\leftrightarrow D_{5/2}(m_{j}=-3/2)$.
The strong suppression of the red sideband transition  indicates near
ground state cooling for all the motional modes. The dash lines denote the position of the cooling resonant frequencies. (c) The average phonon numbers for
the 12 motional modes of the 4-ion chain, thermometry for each motional
mode is estimated using the sideband asymmetry and corresponding correction
factor (see supplement).}
\label{fig:figure4}
\end{figure}

We perform proof of principle experiment on a single ion to study  the optimal
condition of the parallel-EIT cooling
 method and make a comparison for the cooling
limit and bandwidth of the parallel-EIT cooling and EIT cooling.  As shown in Fig. \ref{fig:figure3}(a),  we firstly use Doppler cooling to bring down the average phonon number of
the ion to be around 10, then we switch on the cooling beams simultaneously
for a duration $\tau_{\rm parallel-EIT/EIT}$ to cool down  the motional mode.  Finally we apply a  $\SI{50}{\micro\second}$ optical pumping pulse to prepare the ion in the state $S_{1/2}(m_j = -1/2)$ and apply  blue sideband pulses to  measure the sideband spectroscopy of quadrupole transition $S_{1/2}\leftrightarrow D_{5/2}$  at $\SI{729}{\nano\meter}$ for thermometry. The final average phonon number $\bar{n}$ can
be obtained by fitting experimental data points with the analytical
solution of sideband Rabi oscillation \cite{king1998cooling,Blatt2003RevModPhys}.
 In Fig. \ref{fig:figure3}(b), we measure the average phonon number of parallel-EIT cooling
 for different ac Stark shift $\delta_{ac}$. The axial mode to be cooled has frequency $2\pi\times \SI{1.4}{\mega\hertz}$  and  the optimal ac Stark shift for standard EIT cooling is $2\pi\times 1.4 $ MHz. For  parallel-EIT cooling
  method $\delta_{ac}$ is  adjusted by tuning the  $\Omega_r$ and the detuning $\Delta_{g}$ is chosen according to the optimal condition  (\ref{eqn1}). For each $\delta_{ac}$, we compare the cooling result with EIT cooling. The parallel-EIT cooling
  method shows a slight dependence on the mode frequency and a wide cooling bandwidth  with best final phonon number around 0.2, while EIT cooling shows a limited cooling range as expected. We further investigate the cooling bandwidth of our method and EIT cooling by using fixed ac Stark shift and widely tuned mode frequencies, which simulate the mode spectrum of a string of ions. The motional mode frequencies are varied  between $2\pi\times \SI{0.98}{\mega\hertz}$ and  $2\pi\times \SI{3.6}{\mega\hertz}$  and corresponding $\Delta_{g}$ is adjusted according to the optimal cooling requirement. The cooling results in Fig. \ref{fig:figure3}(c) show that the motional modes under different frequencies can always be cooled down to near ground state with best average phonon number around 0.1, while the EIT cooling shows worse cooling limit as the  mode frequency deviates from the fixed ac Stark shift.



We then investigate simultaneous cooling performance of parallel-EIT cooling method by cooling all the modes of a single and two ions.
For 3D cooling  of a single ion,
the $\sigma^{-}$ light, which inducing an ac Stark shift about $2\pi\times \SI{2.0}{\mega\hertz}$, together with the $\pi$ polarized beam with two
frequency components creates optimal cooling for trap frequencies
$\{\omega_{z},\omega_{x},\omega_{y}\}/2\pi=\{0.6,3.63,3.71\}$MHz.
Fig. \ref{fig:figure4}(a) shows the cooling results for the 3 motional
modes by applying a cooling pulse time of 1.5 ms. The average phonon
numbers for all the motional modes are below 0.5. We also apply this
method for cooling a 2-ion chain with trap frequencies $\{\omega_{z},\omega_{x},\omega_{y}\}/2\pi=\{0.8,3.98,4.02\}$
MHz, where 3 frequency components are added to the probe beam. The
cooling results for 2 ions are also shown in Fig. \ref{fig:figure4}(a),
all the 6 motional modes are cooled down to near ground state with
cooling time of $ \SI{1.5}{\milli\second}$ as the single ion case.  Note
that the trap frequency difference for 1 and 2 ions is over $ \SI{3}{\mega\hertz}$,
which is far beyond the cooling bandwidth of EIT cooling as indicated
in Fig. \ref{fig:figure3}(c), nevertheless, parallel-EIT cooling
 method can  cover all the motional modes.


We finally study the multimode cooling performance of our method by using a 4-ion chain. To maintain the ion crystals in
the trap, the trap frequencies are reduced to $\{\omega_{z},\omega_{x},\omega_{y}\}/2\pi=\{0.6,1.706,1.754\}$
MHz. To cover the whole mode spectrum, 5 frequency components are
used for generating the probe beam. With a cooling time of 1.5 ms,
we observe an efficient simultaneous near ground state cooling for all motional modes as indicated in Fig. \ref{fig:figure4}(b) and more accurate average phonon numbers are shown in Fig. \ref{fig:figure4}(c). The performance of parallel-EIT cooling method is found to be the same for various ion numbers under the assumption that the probe beam has enough power for each frequency components. This experiment proves the feasibility for cooling all motional modes of a large  ion chain.  Howerver, we note that  the cooling bandwidth is restricted by the Doppler cooling limit in practice since our method works near the LD regime. Therefore we need to apply sub-Doppler cooling to bring the axial motional modes with low frequencies  to LD regime before implementing  our method to  a large-size ion string (see supplement).


In summary, we proposed a new  cooling method with arbitrarily tunable cooling bandwidth. Experimentally we proved that the
new cooling method is  efficient  for the simultaneous cooling of
multiple ions by using up to 4 $^{40}$Ca$^{+}$ ions and thoeretically showed the feasibility for simultaneous cooling for a large number of atoms and ions. Since realization of our method only requires  modification on  the probe beam compared to the standard  EIT cooling, our scheme works as long as EIT works and therefore it is  applicable for  ions with hyperfine level structures, such as $^{171}$Yb$^{+}$
ions.

We thank  Kihwan Kim, Yiheng Lin for their helpful discussion. We acknowledge the use of the Quantum Toolbox in PYTHON (QuTiP) \cite{qutip}. This work
is supported by the National Basic Research Program of China (Grant
No. 2016YFA0301903), the National Natural Science Foundation of China
(Grant Nos. 12004430, 11904402,12074433, 12174447, 12174448, 61632021).

\bibliography{arxiv}

\onecolumngrid
\newpage\clearpage{}

\section*{Supplement}

\global\long\def\thefigure{S.\arabic{figure}}%
 \setcounter{figure}{0}
\global\long\def\theequation{S.\arabic{equation}}%
 \setcounter{equation}{0}
\global\long\def\thetable{S.\arabic{table}}%
 \setcounter{table}{0}

\subsection{Theoretical model of parallel-EIT cooling}

\begin{figure}[htbp]
	\centering\includegraphics[width=5cm]{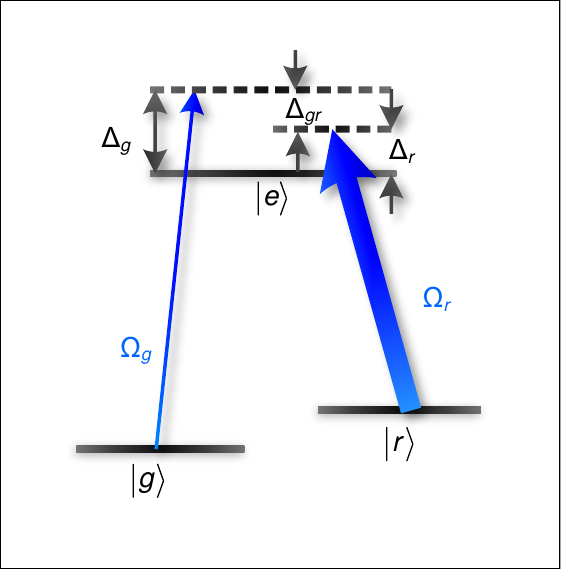}
		\caption{The level diagram of the ion. The ion has a dissipative excited state
			$\left|e\right\rangle $ and two ground states $\left|g\right\rangle $
			and $\left|r\right\rangle $, transitions $\left|e\right\rangle \leftrightarrow\left|g\right\rangle $
			and $\left|e\right\rangle \leftrightarrow\left|r\right\rangle $ are
			driven by two external laser fields.}
		\label{figS1-1}
\end{figure}

We consider that an ion of mass $M$ is confined in
	a harmonic trap with trap frequency $\omega$. The level configuration
	of the trapped ion can be treated as a three-level atom (see Fig.
	\ref{figS1-1}), with a dissipative excited state $\left|e\right\rangle $
	and two ground states $\left|g\right\rangle $ and $\left|r\right\rangle $,
	whose level frequencies are $\omega_{e}$, $\omega_{g}$ and $\omega_{r}$,
	respectively. States $\left|e\right\rangle $ and $\left|g\right\rangle $
	couple to a cooling laser with frequency $\omega_{L1}$ and Rabi frequency
	$\Omega_{g}$, while states $\left|e\right\rangle $ and $\left|r\right\rangle $
	couple to a driving laser with frequency $\omega_{L2}$ and Rabi frequency
	$\Omega_{r}$.

The Hamiltonian of the whole system takes the form
	\begin{equation}
		H=H_{0}+V_{ge}+V_{re},
	\end{equation}
	where $H_{0}$ is the free Hamiltonian of atomic degrees of freedom
	(d.o.f) and motional d.o.f, $V_{ge}$ ($V_{re}$) is the coupling
	between states $\left|g\right\rangle $ and $\left|e\right\rangle $
	($\left|e\right\rangle $ and $\left|r\right\rangle $), (we have set $\hbar =1$ )

	\begin{eqnarray}
		H_{0} & = & -\Delta_{g}\left|e\right\rangle \left\langle e\right|-\Delta_{gr}\left|r\right\rangle \left\langle r\right|+\omega b^{\dagger}b,\nonumber \\
		V_{ge} & = & \frac{\Omega_{g}}{2}\left|e\right\rangle \left\langle g\right|\textrm{e}^{ik_{L1}x\cos\varphi_{g}}+\textrm{h.c.},\nonumber \\
		V_{re} & = & \frac{\Omega_{r}}{2}\left|e\right\rangle \left\langle r\right|\textrm{e}^{ik_{L2}x\cos\varphi_{r}}+\textrm{h.c.}.\label{eq1}
	\end{eqnarray}
	Here $b$($b^{\dagger}$) is the annihilation (creation) operator
	of the ion's vibrational state; the detunings are defined by
	\begin{eqnarray}
		\Delta_{g} & = & \omega_{L1}-\left(\omega_{e}-\omega_{g}\right),\nonumber \\
		\Delta_{r} & = & \omega_{L2}-\left(\omega_{e}-\omega_{r}\right),\nonumber \\
		\Delta_{gr} & = & \Delta_{g}-\Delta_{r};\label{eq2}
	\end{eqnarray}
	the wave numbers are $k_{L1}=\dfrac{\omega_{L1}}{c},k_{L2}=\dfrac{\omega_{L2}}{c}$;
	and $\varphi_{g}$ $\left(\varphi_{r}\right)$ denote the angle between
	the motional axis and cooling (driving) laser.

In the Lamb-Dicke (LD) regime, $V_{ge}$ and $V_{re}$
	can be expanded to $V_{ge}=V_{ge}^{0}+V_{ge}^{1}$ and $V_{re}=V_{re}^{0}+V_{re}^{1}$,
	which are
	\begin{eqnarray}
		V_{ge}^{0} & = & \frac{\Omega_{g}}{2}\left|e\right\rangle \left\langle g\right|+\textrm{h.c.},\nonumber \\
		V_{re}^{0} & = & \frac{\Omega_{r}}{2}\left|e\right\rangle \left\langle r\right|+\textrm{h.c.},\nonumber \\
		V_{ge}^{1} & = & i\eta_{g}\cos\varphi_{g}\frac{\Omega_{g}}{2}\left|e\right\rangle \left\langle g\right|\left(b^{\dagger}+b\right)+\textrm{h.c.,}\nonumber \\
		V_{re}^{1} & = & i\eta_{r}\cos\varphi_{r}\frac{\Omega_{r}}{2}\left|e\right\rangle \left\langle r\right|\left(b^{\dagger}+b\right)+\textrm{h.c.},\label{eq3}
	\end{eqnarray}
	where the superscript indicates the order in LD parameters, and

	\begin{equation}
		\eta_{g}=k_{L1}\sqrt{\frac{1}{2M\omega}},\eta_{r}=k_{L2}\sqrt{\frac{1}{2M\omega}}.\label{eq4}
	\end{equation}

Then the Hamiltonian in LD regime reads
	\begin{equation}
		H_{\textrm{LD}}=H_{0}+V_{ge}^{0}+V_{re}^{0}+V_{ge}^{1}+V_{re}^{1}.
	\end{equation}
	By considering the dissipations, the master equation of the system
	is

	\begin{equation}
	\frac{d}{dt}\rho=  -i\left[H,\rho\right]+\sum_{j=g,r}\frac{\gamma_{j}}{2}(2\left|j\right\rangle \left\langle e\right|\widetilde{\rho}\left|e\right\rangle \left\langle j\right|-\left|e\right\rangle \left\langle e\right|\rho
		 -\rho\left|e\right\rangle \left\langle e\right|),
		\label{eq5}
	\end{equation}
	where
	\begin{equation}
	\ket{j} \expval{e|\widetilde{\rho}|e} \bra{j}=  \frac{1}{2}\int_{-1}^{+1}\textrm{d}\left(\cos\theta_{j}\right)\mathcal{N}_{j}\left(\cos\theta_{j}\right)\left|j\right\rangle \left\langle e\right|
		 \textrm{e}^{ik_{ej}x}\rho\textrm{e}^{-ik_{ej}x}\left|e\right\rangle \left\langle j\right|,
		\label{eq6}
	\end{equation}
	with $\mathcal{N}_{j}\left(\cos\theta_{j}\right)$ being the angular
	distribution, and $\gamma_{j}\left(j=g,r\right)$ denotes the spontaneous
	dissipation rate from state $\left|e\right\rangle $ to $\left|j\right\rangle $.

\begin{figure*}[htbp]
	\centering \includegraphics[width=12cm]{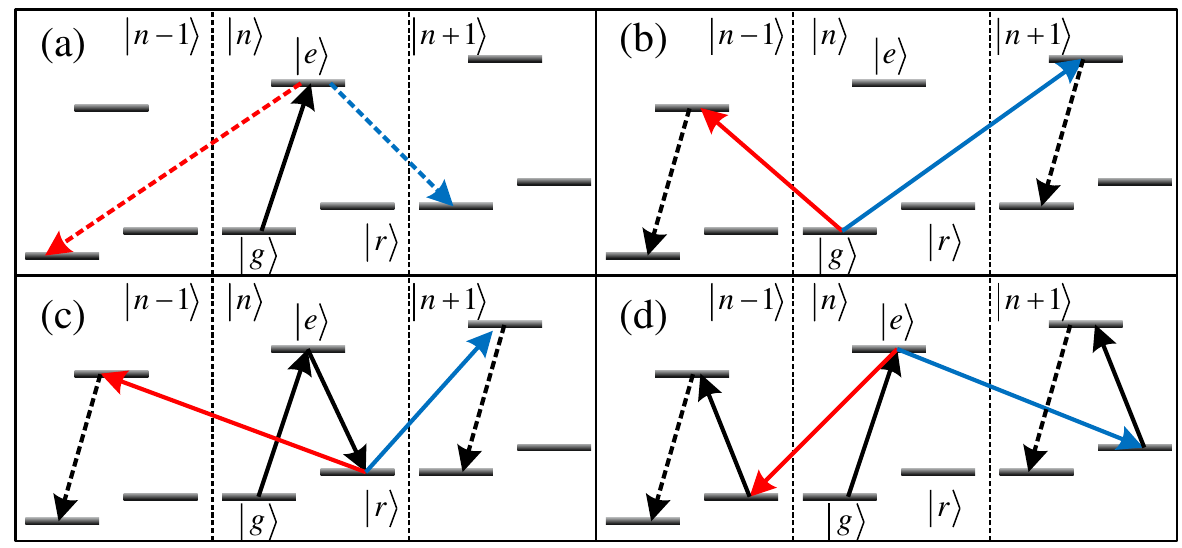}
		\caption{The possible cooling and heating transitions. (a)-(d) correspond to
			transitions (a)-(d) in Eqs.(\ref{eq7})}
		\label{figS1-2}
\end{figure*}

Here, we concern the regime that $\Omega_{g}\ll\Omega_{r}$,
	so that the atomic steady population is almost at state $\left|g\right\rangle $.
	In LD regime, from the steady state of $n$th phonon $\left|g,n\right\rangle $,
	possible heating (cooling) transitions are

	\begin{eqnarray}
		(\textrm{a})\qquad &  & \left|g\right\rangle \left|n\right\rangle \xrightarrow{V_{ge}^{0}}\left|e\right\rangle \left|n\right\rangle \xrightarrow{\textrm{dissipation}}\left|g\right\rangle \left|n\pm1\right\rangle \nonumber \\
		(\textrm{b})\qquad &  & \left|g\right\rangle \left|n\right\rangle \xrightarrow{V_{ge}^{1}}\left|e\right\rangle \left|n\pm1\right\rangle \xrightarrow{\textrm{dissipation}}\left|g\right\rangle \left|n\pm1\right\rangle \nonumber \\
		(\textrm{c})\qquad &  & \left|g\right\rangle \left|n\right\rangle \xrightarrow{V_{ge}^{0}}\left|e\right\rangle \left|n\right\rangle \xrightarrow{V_{re}^{0}}\left|r\right\rangle \left|n\right\rangle \xrightarrow{V_{re}^{1}}\left|e\right\rangle \left|n\pm1\right\rangle 
		 \xrightarrow{\textrm{dissipation}}\left|g\right\rangle \left|n\pm1\right\rangle \nonumber \\
		(\textrm{d})\qquad &  & \left|g\right\rangle \left|n\right\rangle \xrightarrow{V_{ge}^{0}}\left|e\right\rangle \left|n\right\rangle \xrightarrow{V_{re}^{1}}\left|r\right\rangle \left|n\pm1\right\rangle \xrightarrow{V_{re}^{0}} \left|e\right\rangle \left|n\pm1\right\rangle \xrightarrow{\textrm{dissipation}}\left|g\right\rangle \left|n\pm1\right\rangle 
		\label{eq7}
	\end{eqnarray}
	as it is depicted in Fig. \ref{figS1-2}(a)-(d), respectively.

The cooling dynamics can be described by the rate
	equation for mean phonon number
	\begin{equation}
		\frac{d}{dt}\left\langle n\right\rangle =-\left(A_{-}-A_{+}\right)\left\langle n\right\rangle +A_{+}.\label{eq8}
	\end{equation}
	where $A_{+}$ and $A_{-}$ are the heating and cooling transition
	rates. For $A_{-}>A_{+}$, the solution of the rate equation is

	\begin{equation}
		\left\langle n\left(t\right)\right\rangle =n_{ss}+\left(n_{0}-n_{ss}\right)\textrm{e}^{-Wt},\label{eq9}
	\end{equation}
	where $n_{0}$ is the initial phonon number, $W=A_{-}-A_{+}$ is the
	cooling rate, and $n_{ss}=\frac{A_{+}}{W}$ is the steady mean phonon
	number.

In weak sideband coupling regime, we can obtain $A_{\pm}$
by calculating all possible heating and cooling transitions \cite{reve1,reve2,reve3} (\ref{eq7}), yielding
	\begin{eqnarray}
		A_{-} & = & \alpha\gamma\left|\mathcal{T}_{s}\right|^{2}+\gamma\left|\mathcal{T}_{1,-}+\mathcal{T}_{2,-}+\mathcal{T}_{3,-}\right|^{2},\nonumber \\
		A_{+} & = & \alpha\gamma\left|\mathcal{T}_{s}\right|^{2}+\gamma\left|\mathcal{T}_{1,+}+\mathcal{T}_{2,+}+\mathcal{T}_{3,+}\right|^{2}.\label{eq10}
	\end{eqnarray}
The first term of $A_{-}$$\left(A_{+}\right)$ denotes the cooling
(heating) process caused by diffusion process (see Fig. \ref{figS1-1}(a)),
and the second term corresponds to cooling (heating) process by three
cooling (heating) sideband transitions (see Fig. \ref{figS1-2} (b)-(d)). Here,
	\begin{eqnarray}
		\mathcal{T}_{s} & = & \frac{\eta_{g}\Omega_{g}}{2}\frac{\Delta_{gr}}{f\left(0\right)},\nonumber \\
		\mathcal{T}_{1,\pm} & = & -i\eta_{g}\cos\varphi_{g}\frac{\Omega_{g}}{2}\frac{\mp\omega+\Delta_{gr}}{f\left(\mp\omega\right)},\nonumber \\
		\mathcal{T}_{2,\pm} & = & -i\eta_{r}\cos\varphi_{r}\frac{\Omega_{r}\Omega_{g}}{4}\frac{\frac{\Omega_{r}}{2}}{f\left(0\right)}\frac{\mp\omega+\Delta_{gr}}{f\left(\mp\omega\right)},\nonumber \\
		\mathcal{T}_{3,\pm} & = & i\eta_{r}\cos\varphi_{r}\frac{\Omega_{r}\Omega_{g}}{4}\dfrac{\Delta_{gr}}{f\left(0\right)}\frac{\frac{\Omega_{r}}{2}}{f\left(\mp\omega\right)}.\label{eq11}
	\end{eqnarray}
	with
	\begin{equation}
		f\left(x\right)=\left(\Delta_{g}+x\right)\left(\Delta_{gr}+x\right)-\dfrac{\Omega_{r}^{2}}{4}+i\frac{\left(\Delta_{gr}+x\right)\gamma}{2},\label{eq12}
	\end{equation}
	and $\alpha=\intop_{-1}^{+1}\textrm{d}\left(\cos\theta_{j}\right)\cos^{2}\theta_{j}\mathcal{N}_{j}\left(\cos\theta_{j}\right)$.

Efficient cooling is achieved by enhancing cooling
	transition rate $A_{-}$ while suppressing $A_{+}$. For a sufficiently
	strong coupling $\Omega_{r}$, cooling transition rate $A_{-}$ reaches
	its maximum value at
	\begin{equation}
		\textrm{Re}f\left(\nu\right)=\left(\Delta_{g}+\omega\right)\left(\Delta_{gr}+\omega\right)-\dfrac{\Omega_{r}^{2}}{4}=0,\label{eq13}
	\end{equation}
	where the denominator of $A_{-}$ is minimum. This can be understood
	in dressed state representation. As $\Omega_{g}\ll\Omega_{r}$, dressed
	states $\left|+\right\rangle $ and $\left|-\right\rangle $ are created
	by driving beam, which are

	\begin{eqnarray}
		\left|+\right\rangle  & = & \sin\phi\left|e\right\rangle -\cos\phi\left|r\right\rangle ,\nonumber \\
		\left|-\right\rangle  & = & \cos\phi\left|e\right\rangle +\sin\phi\left|r\right\rangle ,\label{eq14}
	\end{eqnarray}
	with $\phi=\frac{1}{2}\arctan\frac{\left|\Omega_{r}\right|}{\Delta_{r}}$.
	The corresponding energies of $\left|+\right\rangle $ and $\left|-\right\rangle $
	are
	\begin{equation}
		E_{\pm}=\frac{1}{2}\left(-\Delta_{r}\pm\sqrt{\Omega_{r}^{2}+\Delta_{r}^{2}}\right),
	\end{equation}
	By tuning the cooling laser in resonance with red sideband transition
	of the narrow-line dressed state $\left|+\right\rangle $, i.e., $\left|g\right\rangle \left|n\right\rangle \longrightarrow\left|+\right\rangle \left|n-1\right\rangle $,
	detunings should satisfy

	\begin{equation}
		\left|\Delta_{gr}+\omega\right|=\delta_{ac},\label{eq16}
	\end{equation}
	One can verify that the red sideband resonance condition is precisely
	the same as optimal condition (\ref{eq13}).

Assuming that $\left|\Delta_{g}\right|\gg\omega,\left|\Delta_{gr}\right|,\delta_{ac}$,
	we obtain
	\begin{eqnarray*}
		A_{-} & \simeq & \eta_{1}^{2}\frac{\Omega_{g}^{2}}{\gamma}\\
		A_{+} & \simeq & \eta_{2}^{2}\left(\frac{\gamma\Omega_{g}^{2}}{16\Delta_{g}^{2}}\right)
	\end{eqnarray*}
	with the effective Lamb-Dicke parameters $\eta_{1}$ and $\eta_{2}$
	defined by

	\begin{eqnarray*}
		\eta_{1}^{2} & = & \eta^{2},\\
		\eta_{2}^{2} & = & \left\{ \begin{array}{ll}
			\eta^{2}\left(1+4\alpha\left(1-\frac{\delta_{ac}}{\omega}\right)^{2}\right) & \Delta_{g}>0,\\
			\eta^{2}\left(1+4\alpha\left(1+\frac{\delta_{ac}}{\omega}\right)^{2}\right) & \Delta_{g}<0.
		\end{array}\right.
	\end{eqnarray*}
Here, we have defined that $\eta_{g}\approx\eta_{r}=\frac{\eta}{2}$,
	and assumed $\varphi_{g}=0,\varphi_{r}=\pi$ for simplicity. The corresponding
	final mean phonon number is
	\[
	n_{st}=Rn_{\textrm{EIT}},
	\]
	where $R$ is the ratio of parallel-EIT cooling's $n_{\textrm{st}}$ to EIT cooling
	$n_{\textrm{EIT}}=\frac{\gamma^{2}}{16\Delta_{g}^{2}}$, that is

	\begin{eqnarray*}
		R & = & \left\{ \begin{array}{ll}
			1+4\alpha\left(1-\frac{\delta_{ac}}{\omega}\right)^{2} & \Delta_{g}>0,\\
			1+4\alpha\left(1+\frac{\delta_{ac}}{\omega}\right)^{2} & \Delta_{g}<0.
		\end{array}\right.
	\end{eqnarray*}

\subsection{Multimode cooling via parallel-EIT cooling}

To theoretical analysis of multimode cooling via parallel-EIT cooling,
	we first consider the simplest case that cooling of a single ion's
	two-dimensional (2D) motion with two separated motional modes. We
	apply two probe laser beams and one driving beam. The interaction
	Hamiltonian takes the form
	\[
	H=H_{0}+V.
	\]
	Here, $H_{0}$ is the free Hamiltonian for atomic degree of freedom
	(d.o.f) and motional d.o.f
	\[
	H_{0}=\omega_{z}a_{z}^{\dagger}a_{z}+\omega_{x}a_{x}^{\dagger}a_{x}-\Delta_{g1}\left|e\right\rangle \left\langle e\right|-\Delta_{g1r}\left|r\right\rangle \left\langle r\right|,
	\]
	where $\Delta_{g1}$ is the detuning of the first probe laser, $\Delta_{g1r}=\Delta_{g1}-\Delta_{r}$,
	and $\omega_{z}$ and $\omega_{x}$ are the trap frequencies of axial
	and radial modes.

The laser-ion interaction term $V$ is

\begin{eqnarray}
	&V  &=  \frac{\Omega_{g1}}{2}\left|e\right\rangle \left\langle g\right|\exp\left[i\sum_{k=x,z}\eta_{gk}\left(a_{k}+a_{k}^{\dag}\right)\cos\theta_{k1}\right]		   +\frac{\Omega_{g2}}{2}\left|e\right\rangle \left\langle g\right|\exp\left[i\sum_{k=x,z}\eta_{gk}\left(a_{k}+a_{k}^{\dag}\right)\cos\theta_{k2}   +i\Delta_{g12}t\right] \\ \nonumber
	&  &	 +\frac{\Omega_{r}}{2}\left|e\right\rangle \left\langle r\right|+\textrm{h.c.},
\end{eqnarray}
where $\Delta_{g12}=\Delta_{g1}-\Delta_{g2}$ is the frequency difference
between the detunings of two probe lasers, $\Delta_{g2}$ is the detuning of the second probe laser; $\eta_{gk}=k_{L}\sqrt{\frac{1}{2M\omega_{k}}}\left(k=x,z\right)$;
$\theta_{kl}$ is the angle between the $l$-th probe laser and $k-$axis.
For simplicity, we assume that the driving laser is perpendicular
to the two axes. In dressed state basis and the interaction picture,
the Hamiltonian takes the form

\begin{eqnarray}
	 	H  = \frac{\Omega_{+1}}{2}\left|+\right\rangle \left\langle g\right|\exp\left[i\sum_{k=x,z}\eta_{gk}\left(a_{k}e^{-i\omega_{k}t}+a_{k}^{\dag}e^{i\omega_{k}t}\right)\cos\theta_{k1}\right]
	 e^{i\left(-\Delta_{g1r}+\delta_{ac}\right)t}+\textrm{h.c.} \\ \nonumber
		 +\frac{\Omega_{+2}}{2}\left|+\right\rangle \left\langle g\right|\exp\left[i\sum_{k=x,z}\eta_{gk}\left(a_{k}e^{-i\omega_{k}t}+a_{k}^{\dag}e^{i\omega_{k}t}\right)\cos\theta_{k2}\right]
	e^{i\left(-\Delta_{g2r}+\delta_{ac}\right)t}+\textrm{h.c.},
	\end{eqnarray}
where we have neglected the far-off resonant state $\left|-\right\rangle $.

\begin{figure}[htbp]
\noindent
\begin{minipage}[c]{1\linewidth}%
\centering \includegraphics[width=0.45\textwidth]{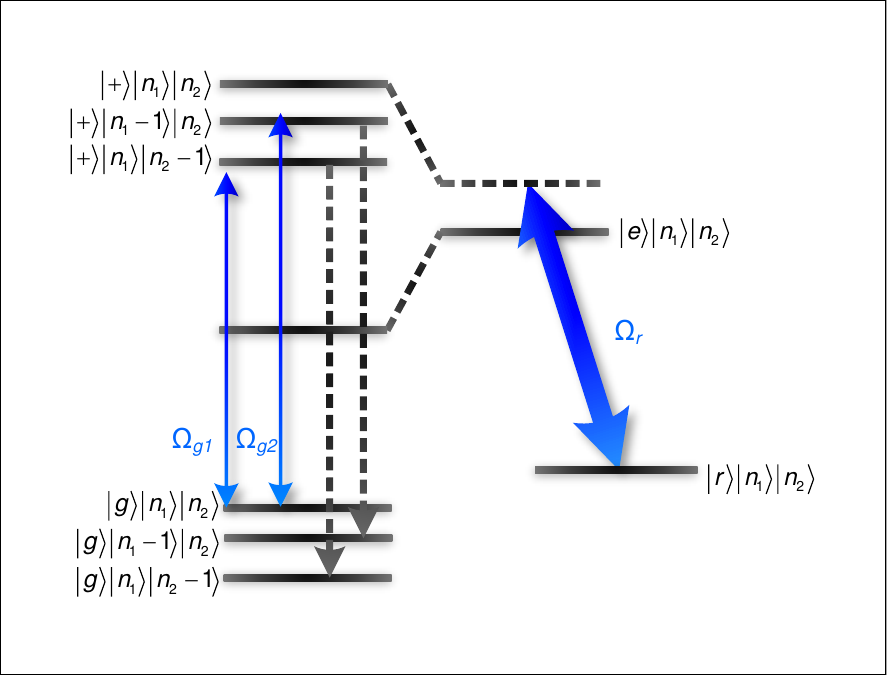}
\includegraphics[width=0.45\textwidth]{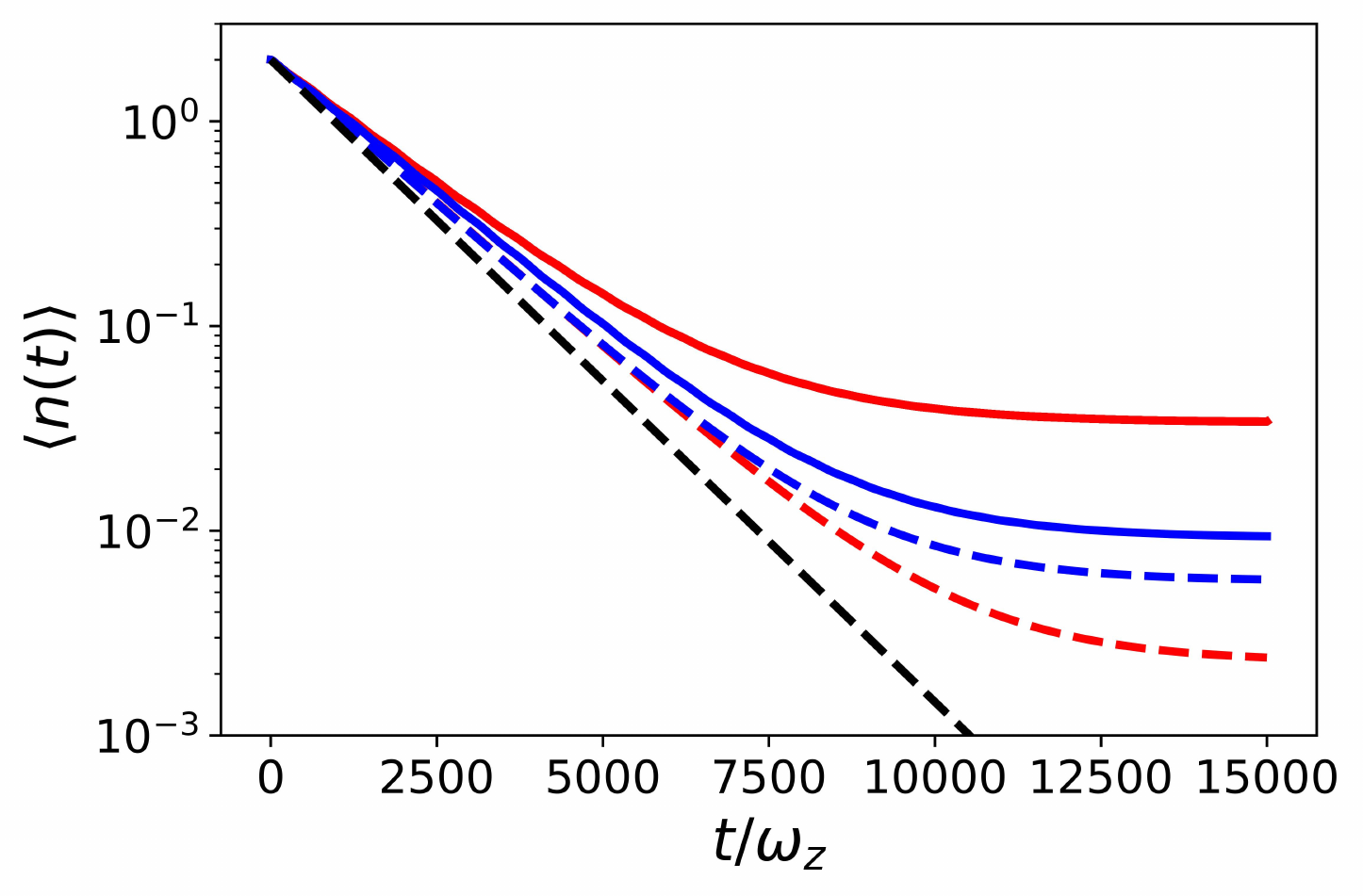}
		
\caption{(a) Cooling principle for simultaneously cooling 2D motional modes
of a single ion. Two probe laser beams and one driving beam create
cooling resonances for the widely separated axial mode ($\omega_{z}$
) and radial mode ($\omega_{x}$). (b) The average photon number $\left\langle n\left(t\right)\right\rangle $ as a function of $t$ for multimode parallel-EIT cooling. The red and blue solid lines	are simultaneously cooling dynamics for axial mode and radial mode,
respectively. The red and blue dashed lines correpond to the cooling
dynamics of single mode cooling at the same parameters. The black
dashed line is the analytical prediction. The simulation parameters
are $\omega_{x}=10\omega_{z}$, $\delta_{ac}=2\omega_{z}$, $\Delta_{r}=330\omega_{z}$,	$\gamma=20\omega_{z}$, $\Omega_{g1}=\omega_{z}$, $\Delta_{g1}=\left(331+0.005\right)\omega_{z}$,
$\Omega_{g2}=3.16\omega_{z}$, $\Delta_{g2}=\left(322-0.005\right)\omega_{z}$,
$\eta_{gz}=0.17$, $\theta_{k1}=\theta_{k2}=\pi/4$.}
\label{figS2-1}
\end{minipage}
\end{figure}

In LD regime, by tuning that
\begin{equation}\label{S21}
\Delta_{g1r}+\omega_{z}=\Delta_{g2r}+\omega_{x}=\delta_{ac}
\end{equation}
and neglect off-resonant transitions, we obtain the effective Hamiltonian

\begin{equation}
		H_{\textrm{eff}}  =  i\eta_{gz}\frac{\Omega_{+1}}{2}\left|+\right\rangle \left\langle g\right|a_{z}+\textrm{h.c.}	 +i\eta_{gx}\frac{\Omega_{+2}}{2}\left|+\right\rangle \left\langle g\right|a_{x}+\textrm{h.c.}.
\end{equation}
In weak coupling regime, either mode is cooled independently from
the other (Fig. \ref{figS2-1} (a)). However, due to the multi-phonon
resonances, the whole system may evolve into some dark states, e.g.
$\alpha_{1}\left|g,2_{x},0_{z}\right\rangle +\alpha_{2}\left|g,1_{x},1_{z}\right\rangle +\alpha_{3}\left|g,0_{x},2_{z}\right\rangle $,
which decouples from the $H_{\textrm{eff}}$. To avoid the dark resonance,
we can set the detunings such that conditions (\ref{S21}) are slightly mismatched,
i.e.,	\[
	\Delta_{g1r}+\omega_{z}+\varepsilon=\Delta_{g2r}+\omega_{x}-\varepsilon=\delta_{ac}.
	\]
	where $\left|\varepsilon\right|\ll\omega_{x,z}$. This way, the 2D
	motion can be cooled at the same time.

In Fig. \ref{figS2-1} (b), we numerically simulate
the simultaneous 2D cooling. To highlight the advantage of our scheme,
we consider an extreme case that $\omega_{x}=10\omega_{z}$, where
standard EIT cooling have to sequentially cool the two modes. We also
plot parallel-EIT cooling of single mode for comparison. The simulation shows that
the simultaneous 2D cooling has the same cooling rates as those of
single mode cooling. 

We further consider that an $N$-ion chain with $3N$ collective modes are globally coupled by $m$ probe lasers and one driving laser. The Hamiltonian takes the form

\begin{equation}
H  =  \sum_{j=1}^{N}\sum_{l=1}^{m}\frac{\Omega_{+l}}{2}\left|+\right\rangle _{jj}\left\langle g\right|\exp\left[i\sum_{k=1}^{3N}\eta_{jk}\left(a_{k}e^{-i\omega_{k}t}+a_{k}^{\dag}e^{i\omega_{k}t}\right)\right]e^{i\left(-\Delta_{glr}+\delta_{ac}\right)t}+\textrm{h.c.},
\end{equation}
where the indices $j,k,l$ indicate the $j-$th ion, $k-$th collective
mode and $l-$th probe laser, $\eta_{jk}=k_{L}\sqrt{\frac{1}{2M\omega_{k}}}$$b_{jk}$
with $b_{jk}$ being the eigenvectors for the mode $k$. One can tune
the frequencies of probe lasers, such that each probe laser is tuned
to near resonace with the red sideband transition for a range of nearby
modes. As a result, all the motional modes can be cooled near the
ground state. However, each of the modes may be affected by heating effects from
all probe lasers. The heating coefficient of the mode at frequency
$\omega$ is $A_{+}\left(\omega\right)=\sum_{l=1}^{m}A_{+l}\left(\omega\right)$,
where\[
A_{+l}\sim\gamma_{+}\eta^{2}\left(\frac{\Omega_{l+}}{2}\right)^{2}\left[\frac{\alpha\left(\frac{\Delta_{glr}}{\omega}\right)^{2}}{\left(\Delta_{glr}-\delta_{ac}\right)^{2}+\frac{\text{\ensuremath{\gamma}}_{+}^{2}}{4}\left(\frac{\Delta_{glr}}{\omega}\right)^{2}}+\frac{\left(\frac{\Delta_{glr}-\omega}{\omega}\right)^{2}}{\left(\Delta_{glr}-\delta_{ac}-\omega\right)^{2}+\frac{\text{\ensuremath{\gamma}}_{+}^{2}}{4}\left(\frac{\Delta_{glr}-\omega}{\omega}\right)^{2}}\right].
\]
Therefore, the final mean phonon number roughly scales linearly with the number of the probe beams.

\begin{figure}[htbp]
	\centering
	\includegraphics[width=0.45\textwidth]{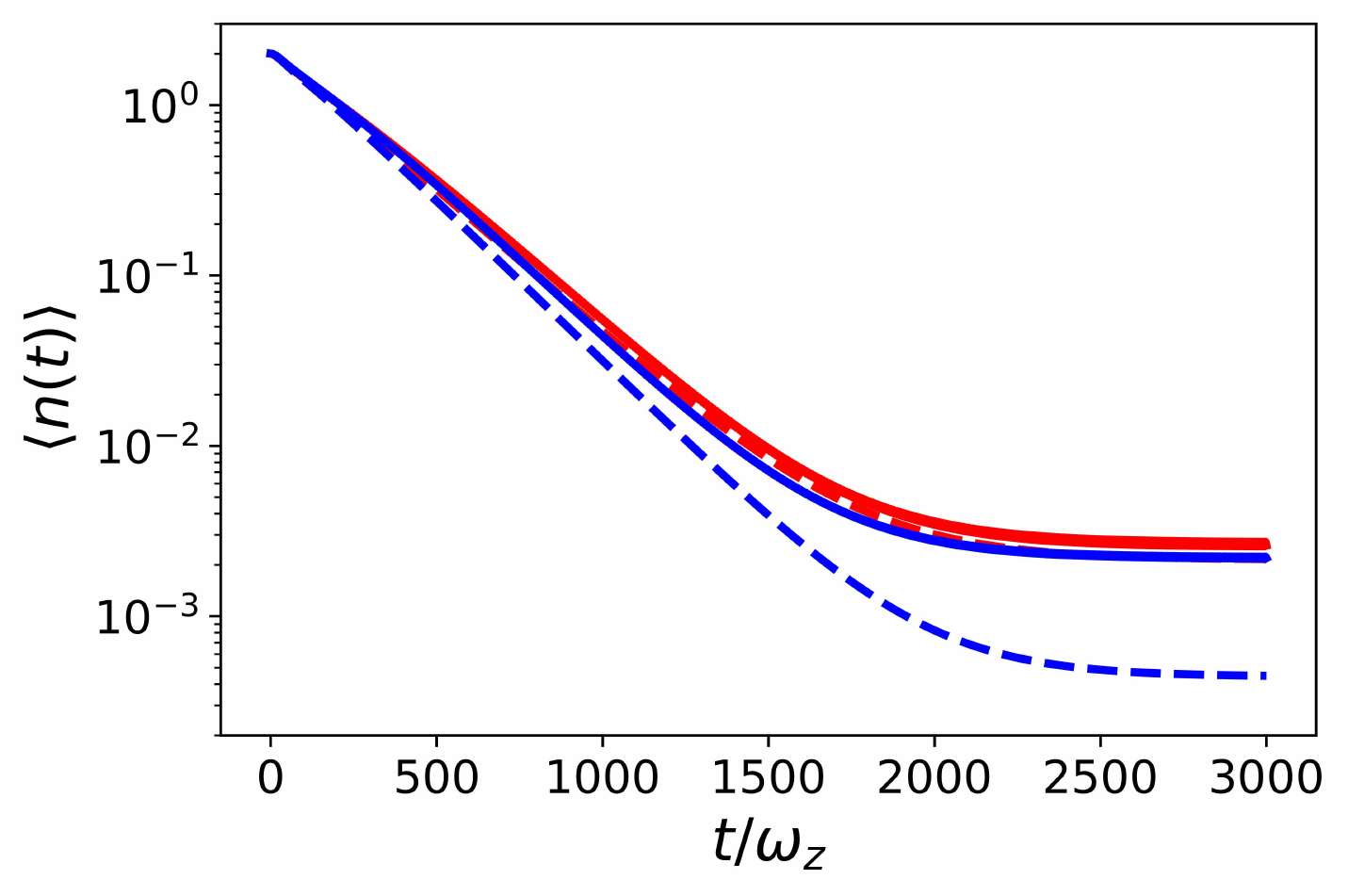}
	\caption{The average photon number $\left\langle n\left(t\right)\right\rangle $
	as a function of $t$ for simultaneous cooling two axial modes of
	a two-ion crystal. The red and blue solid lines stand for center-of-mass
	mode ($\omega_{z}$) and stretch mode ($\sqrt{3}\omega_{z}$). The
	dashed lines correpond to the cooling dynamics of single mode cooling
	at the same parameters. The simulation parameters are $\delta_{ac}=2\omega$,
	$\Delta_{r}=330\omega$, $\gamma=20\omega_{z}$, $\Omega_{g1}=\omega_{z}$,
	$\Delta_{g1}=\left(331+0.005\right)\omega_{z}$, $\Omega_{g2}=1.32\omega_{z}$,
	$\Delta_{g2}=\left(330.27-0.005\right)\omega_{z}$, $\eta_{\text{COM}}=0.24$,
	$\eta_{\text{strech}}=0.18$.}
		\label{figS2-2}
\end{figure}

In Fig. \ref{figS2-2}, we demonstrate the numerical
	simulation of the simultaneous cooling two axial modes of a two-ion
	crystal. The two ions are globally coupled by two probe lasers. For
	both of the probe lasers illuminate the two ions, the cooling rate
	of the ion string is the same as that of cooling a single ion.
	
For cooling all motional modes of a long  ion-string ,  we need to consider the size of  the ion-crystal since our method works in LD regime. Assuming  cooling of a  linear string of 40 ion with transverse trap frequency about 4.5 MHz as in Ref. \cite{reve7}, the maximum axial COM mode  frequency to keep the  linear string is  about 280 kHz according to the theory in Ref. \cite{reve8},  in this case the average phonon number  would be quite high  after Doppler cooling, our method cannot be applied since the  ions are far from LD regime,  therefore the cooling mode frequency  of our method  cannot be arbitrarily  low. Experimentally the cooling bandwidth of our method can be further extended with the help of sub-Doppler cooling method such as Sisyphus cooling.

\subsection{Experimental setup for parallel-EIT cooling }

\begin{figure}

\begin{minipage}[c]{0.45\linewidth}%
	\centering \includegraphics[width=1\textwidth]{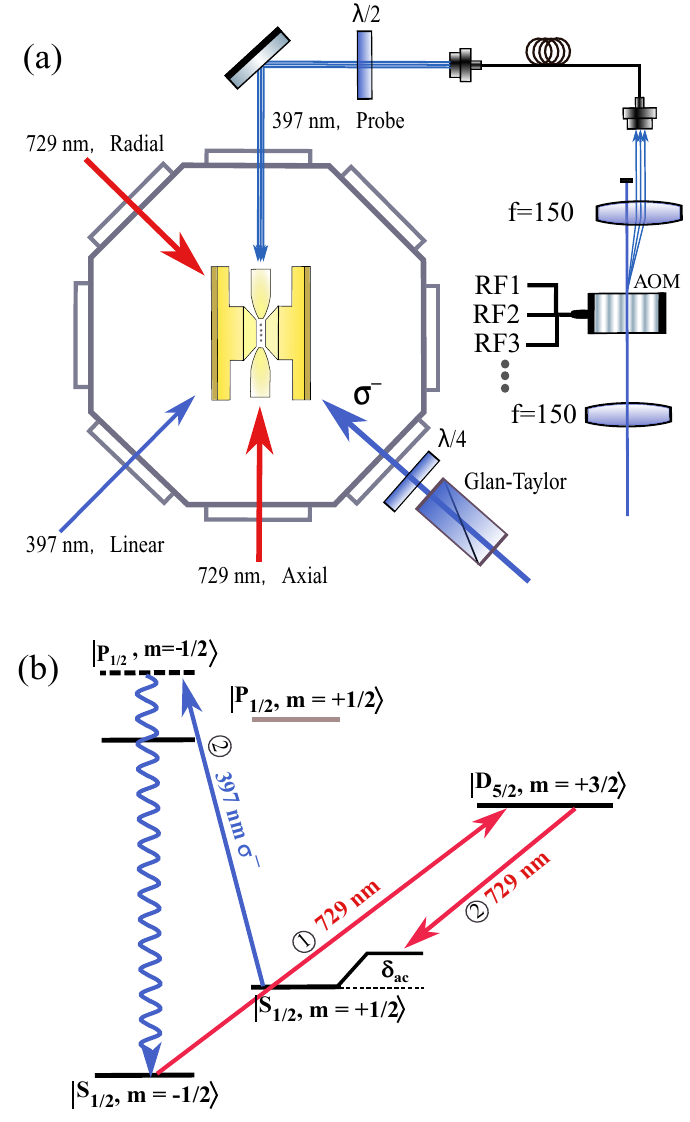} %
\end{minipage}%

\begin{minipage}[c]{0.45\linewidth}%
	\centering \includegraphics[width=1\textwidth]{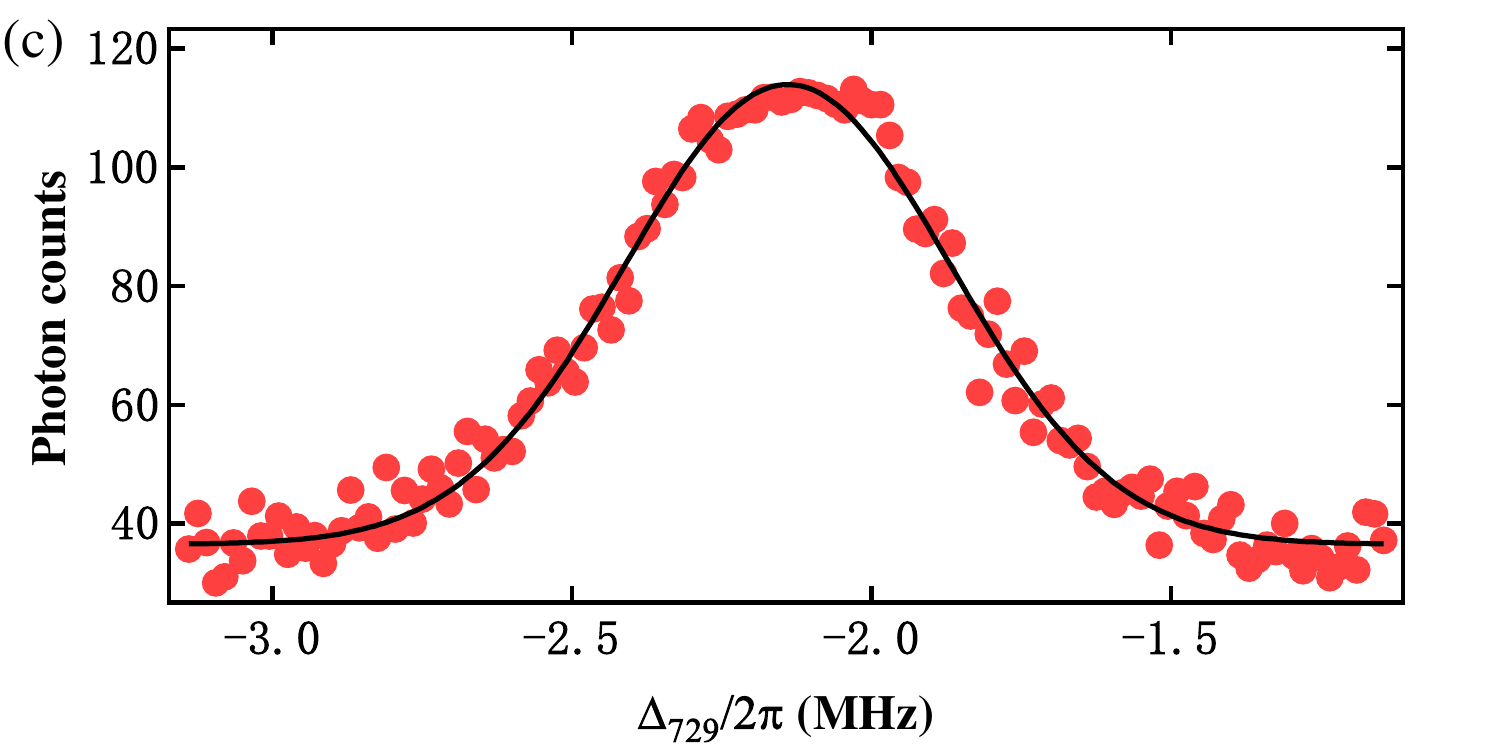} %
\end{minipage}

\caption{(a) Geometric laser configurations for  parallel-EIT cooling  experiments. The coherent
cooling requires both the $\sigma^{-}$ beam and the probe beam, which
is generated by injecting multiple RF frequencies to the single pass
AOM. 397 linear beam is used for the Doppler cooling and fluorescence
detection. The two 729 nm laser beams are used for the thermometry.
(b) The setup for the measurement of the ac Stark shift. One 729 laser
beam stimulates the ion from $\ket{S_{1/2},m_{j}=-1/2}$ to $\ket{D_{5/2},m_{j}=-3/2}$
in the first step, then the driving beam are switched on while scanning
the second 729 nm laser beam in second step, the ac Stark shift can
be obtained by comparing the 729 nm laser frequency with and without
driving beam. (c) shows the measured ac Stark shift to be 2.14 MHz, which is extract from the Gaussian fitting.   }
\label{fig:figs2-1}
\end{figure}

In the coherent cooling method, two 397 nm laser beams are used to
stimulate the transition $S_{1/2}\leftrightarrow P_{1/2}$ of $^{40}$Ca$^{+}$
ions as shown in Fig. \ref{fig:figs2-1}(a). The beam with $\pi$ polarization
goes along the trap axis working as the probe beam, while the $\sigma^{-}$
beam having a $\ang{45}$ with respect to the trap axis is the driving
beam. The wavevector difference $\Delta\vec{k}$ for the two beams
has components along both transverse and axial direction of the ion
chain, therefore this setup can realize simultaneous cooling on all
motional modes. The frequency difference $2\pi\times15.36$ MHz between
the two beams is created by the acousto-optic-modulators (AOM), and
the driving beam has a blue detuning $\Delta_{r}=2\pi\times330$ MHz
from the $\ket{r}\leftrightarrow\ket{e}$ transition. The frequency
of $\pi$ beam in the experiment is determined by using the optimal
condition depending on the motional frequency. For simultaneous cooling
of multiple mode, several RF signals with predetermined frequencies
are combined with a RF frequency combiner/splitter and added to a
single pass AOM on the probe beam path as shown in Fig. \ref{fig:figs2-1}(a).

The ac Stark shift of the driving beam is determined by the spectroscopy
of the quadrupole transition $S_{1/2}\leftrightarrow D_{5/2}$ \cite{reve4}.
As the fact that the driving beam induces same amount of ac Stark
shift for states $\ket{r}$ and $\ket{e}$, the quadrupole transition
frequency difference with and without driving beam is the ac Stark
shift. Experimentally the measurement of $\sigma^{-}$ beam induced
ac Stark shift starts with preparing the ion in the state $\ket{S_{1/2},m_{j}=-1/2}$,
next the 729 nm $\pi$ pulse flips the state to $\ket{D_{5/2},m_{j}=-3/2}$,
finally the driving beam and the 729 nm laser beam addressing the $\ket{D_{5/2},m_{j}=-3/2}\leftrightarrow\ket{S_{1/2},m_{j}=1/2}$
are switched on simultaneously. The following measurement on the D
state population is recorded as a function of the 729 nm laser frequency,
the position with maximum absorption is the resonant frequency of
the transition and the frequency shift from the case without pumping
beam is the ac Stark shift.

Since the motional mode frequency ranges from $2\pi\times0.6-4.0$
MHz, We set the ac Stark shift at $2\pi\times2$ MHz to cover the
cooling bandwidth. The parameters for the probe beam is not calibrated,
we search the optimal working parameters according to the cooling
results. For simultaneous cooling of multiple ions, the probe beam
has several frequency components and their parameters for each component
is adjusted based the cooling results. The cooling pulses include
a Doppler cooling pulse of 1.0 ms and a coherent cooling pulse time
of 1.5 ms.

\subsection{Thermometry thoery for multiple ions}

For 1 and 2 ions, the average phonon number after cooling can be found
by using the motional sensitive Rabi oscillation since the analytical
solution of the blue sideband Rabi oscillation can be found \cite{reve5}
by the laser-atom interaction Hamiltionian. Alternatively the cooling
results for a motional mode can also be measured by using the sideband
spectroscopy, which includes the measurement on the blue and red-sideband.
Under the assumption of thermal distribution, the average phonon number
$\bar{n}$ for a single ion can be extracted by using the sideband
asymmetry $p_{r}/(p_{b}-p_{r})$, where $\bar{n}=p_{r}/(p_{b}-p_{r})$,
$p_{b}$ and $p_{r}$ are the probability of blue and red-sideband
Rabi oscillations \cite{reve6}. For an ion-chain with
more than 2 ions, there is only numerical solutions for the Rabi oscillation.
Instead of fitting to the sideband Rabi oscillation, we use numerical
simulations to find the correction factor for sideband thermometry
method such that the average phonon number is estimated to be product
of the sideband asymmetry and the correction factor.

In our experiment, we chose $\ket{S_{1/2}(m_{j}=-1/2)}(\ket{\downarrow})$
and $\ket{D_{5/2},m_{j}=-3/2}(\ket{\uparrow})$ as the qubit states.
A 729 nm laser beam shining the ions globally with same intensity
is used to drive the quadrupole transition. The detuning $\delta$
is the frequency difference between the 729 nm laser and qubit transition.

\begin{table}
\caption{Mode dependent correction factor}
\label{tab:tableS1} \setlength{\tabcolsep}{5mm}{ %
\begin{tabular}{|c||l||l||l||l|c||l||l||l||l|c||l||l||l||l|c||l||l||l||l|}
\hline
\multicolumn{5}{|c|}{\multirow{2}{*}{Mode}} & \multicolumn{15}{c|}{\multirow{2}{*}{Correction factor}}\tabularnewline
\multicolumn{5}{|c|}{} & \multicolumn{15}{c|}{}\tabularnewline
\cline{6-20} \cline{7-20} \cline{8-20} \cline{9-20} \cline{10-20}
 \cline{11-20} \cline{12-20} \cline{13-20} \cline{14-20} \cline{15-20} \cline{16-20} \cline{17-20} \cline{18-20} \cline{19-20} \cline{20-20}
\multicolumn{5}{|c|}{} & \multicolumn{5}{c|}{\multirow{2}{*}{Y}} & \multicolumn{5}{c|}{\multirow{2}{*}{X}} & \multicolumn{5}{c|}{\multirow{2}{*}{Z}}\tabularnewline
\multicolumn{5}{|c|}{} & \multicolumn{5}{c|}{} & \multicolumn{5}{c|}{} & \multicolumn{5}{c|}{}\tabularnewline
\hline
\multicolumn{5}{|c|}{\multirow{3}{*}{COM}} & \multicolumn{5}{c|}{\multirow{3}{*}{2.10(4)}} & \multicolumn{5}{c|}{\multirow{3}{*}{2.10(4)}} & \multicolumn{5}{c|}{\multirow{3}{*}{2.11(4)}}\tabularnewline
\multicolumn{5}{|c|}{} & \multicolumn{5}{c|}{} & \multicolumn{5}{c|}{} & \multicolumn{5}{c|}{}\tabularnewline
\multicolumn{5}{|c|}{} & \multicolumn{5}{c|}{} & \multicolumn{5}{c|}{} & \multicolumn{5}{c|}{}\tabularnewline
\hline
\multicolumn{5}{|c|}{\multirow{3}{*}{Sym}} & \multicolumn{5}{c|}{\multirow{3}{*}{2.471(6)}} & \multicolumn{5}{c|}{\multirow{3}{*}{2.471(6)}} & \multicolumn{5}{c|}{\multirow{3}{*}{2.363(9)}}\tabularnewline
\multicolumn{5}{|c|}{} & \multicolumn{5}{c|}{} & \multicolumn{5}{c|}{} & \multicolumn{5}{c|}{}\tabularnewline
\multicolumn{5}{|c|}{} & \multicolumn{5}{c|}{} & \multicolumn{5}{c|}{} & \multicolumn{5}{c|}{}\tabularnewline
\hline
\multicolumn{5}{|c|}{\multirow{3}{*}{Asym}} & \multicolumn{5}{c|}{\multirow{3}{*}{2.06(3)}} & \multicolumn{5}{c|}{\multirow{3}{*}{2.06(3)}} & \multicolumn{5}{c|}{\multirow{3}{*}{2.04(3)}}\tabularnewline
\multicolumn{5}{|c|}{} & \multicolumn{5}{c|}{} & \multicolumn{5}{c|}{} & \multicolumn{5}{c|}{}\tabularnewline
\multicolumn{5}{|c|}{} & \multicolumn{5}{c|}{} & \multicolumn{5}{c|}{} & \multicolumn{5}{c|}{}\tabularnewline
\hline
\multicolumn{5}{|c|}{\multirow{3}{*}{ZZ}} & \multicolumn{5}{c|}{\multirow{3}{*}{2.484(8)}} & \multicolumn{5}{c|}{\multirow{3}{*}{2.484(8)}} & \multicolumn{5}{c|}{\multirow{3}{*}{2.463(6)}}\tabularnewline
\multicolumn{5}{|c|}{} & \multicolumn{5}{c|}{} & \multicolumn{5}{c|}{} & \multicolumn{5}{c|}{}\tabularnewline
\multicolumn{5}{|c|}{} & \multicolumn{5}{c|}{} & \multicolumn{5}{c|}{} & \multicolumn{5}{c|}{}\tabularnewline
\hline
\end{tabular}}
\end{table}

In the Lamb-Dicke regime, which is always the case after the  pre-cooling, the laser-atom interaction Hamiltonian can be expressed as
\begin{equation}
H_{I}=\frac{\hbar\Omega}{2}\sum_{j=1}^{N}\sigma_{+j}{\rm exp}\big[i\sum_{k=1}^{3N}\eta_{jk}(a_{k}e^{-i\omega_{k}t}
		+a_{k}^{\dag}e^{i\omega_{k}t})-i(\Delta t-\phi_{j})\big]+h.c.,
\end{equation}
where $k$ and $j$ indicate the colletive mode and ion number respectively.
$\Omega$ is the Rabi frequency, $\eta_{jk}=\frac{2\pi}{\lambda}\sqrt{\frac{\hbar}{2m\omega_{k}}}b_{jk}$
is the Lamb-Dicke parameter, where $m$ is the mass of $^{40}$Ca$^{+}$
ion, $\omega_{k}$ is the mode frequency, $\lambda$ is the wavelength
of 729 nm laser and $b_{jk}$ is the eigenvectors for the mode $k$.
$\sigma$ is the operator for qubit and $a$ ($a^{\dagger}$) is the
annihilation (creation) operator.

When the detuning $\Delta$ is set at the frequency of a motional
mode $\omega_{k}$ ($-\omega_{k}$), the Hamiltionian $H_{I}$ can
be simplified to
\begin{align}
	H_{r,k} & =\frac{i\hbar\Omega}{2}\sum_{j=1}^{N}\sigma_{+j}\eta_{jk}a_{k}+h.c.,\\
	H_{b,k} & =\frac{i\hbar\Omega}{2}\sum_{j=1}^{N}\sigma_{+j}\eta_{jk}a_{k}^{\dagger}+h.c.,
\end{align}
where $H_{r,k}$ and $H_{b,k}$ are Hamiltonians for the red and blue
sideband transitions.

\begin{figure}
\centering %
\begin{minipage}[c]{0.45\linewidth}%
 \centering \includegraphics[width=1\textwidth]{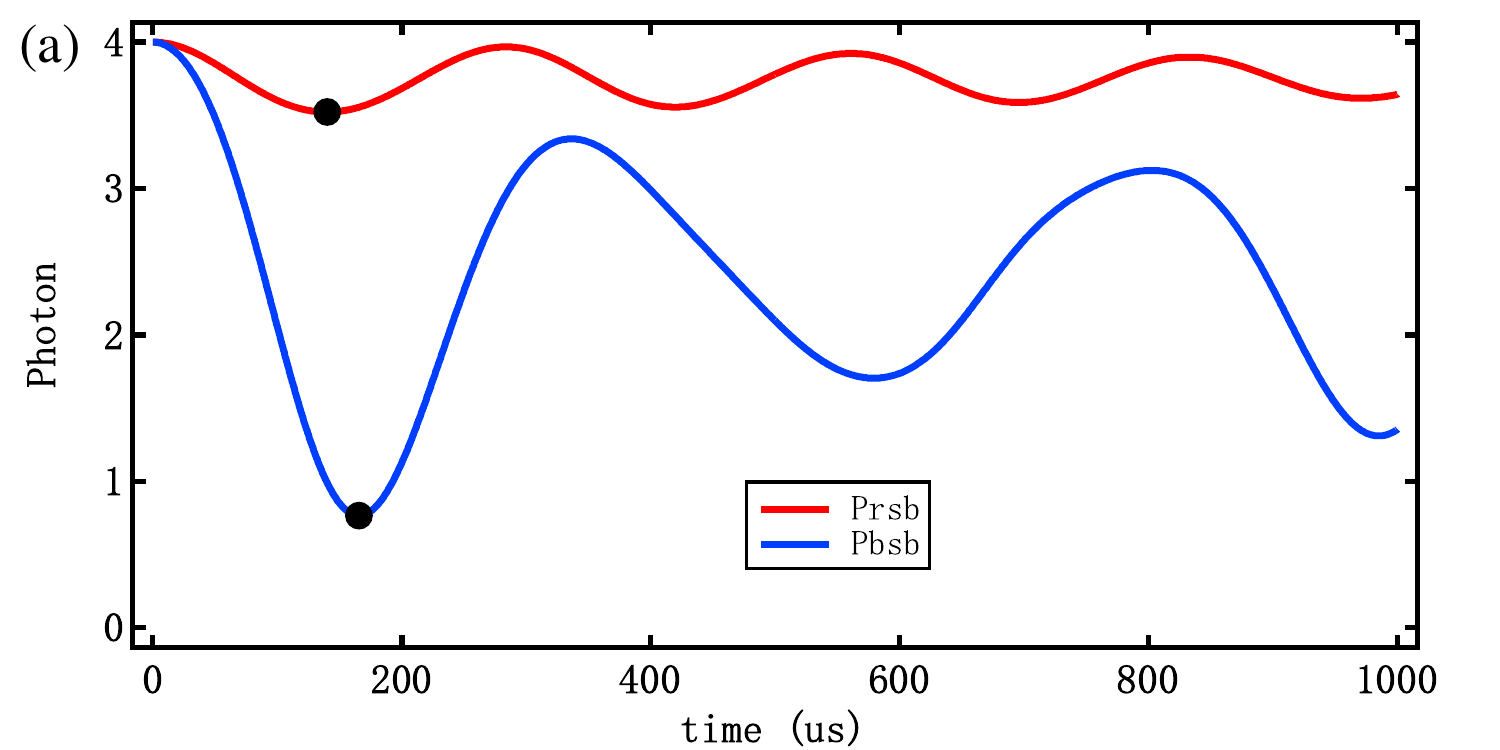} %
\end{minipage}%
\begin{minipage}[c]{0.45\linewidth}%
 \centering \includegraphics[width=1\textwidth]{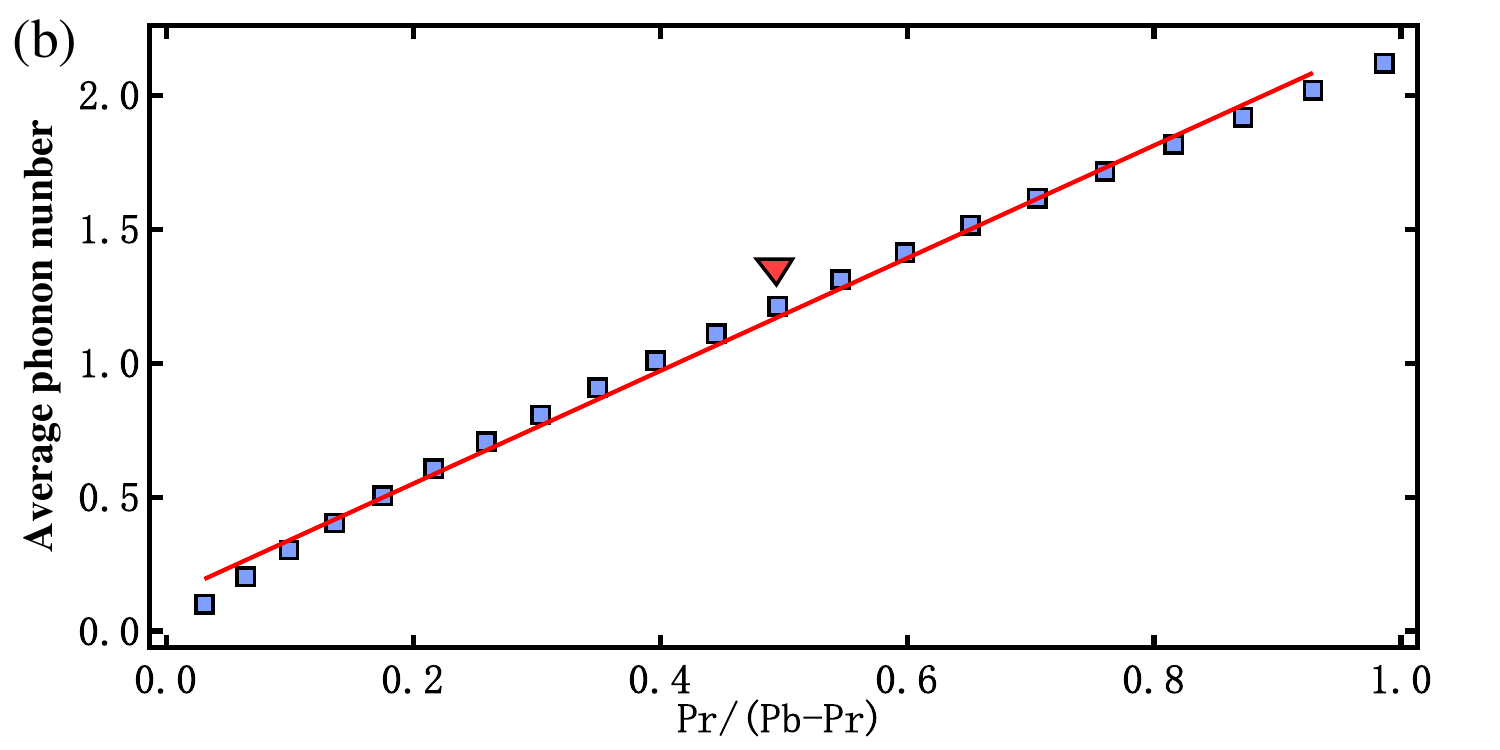} %
\end{minipage}

\begin{minipage}[c]{0.5\linewidth}%
	\centering \includegraphics[width=1\textwidth]{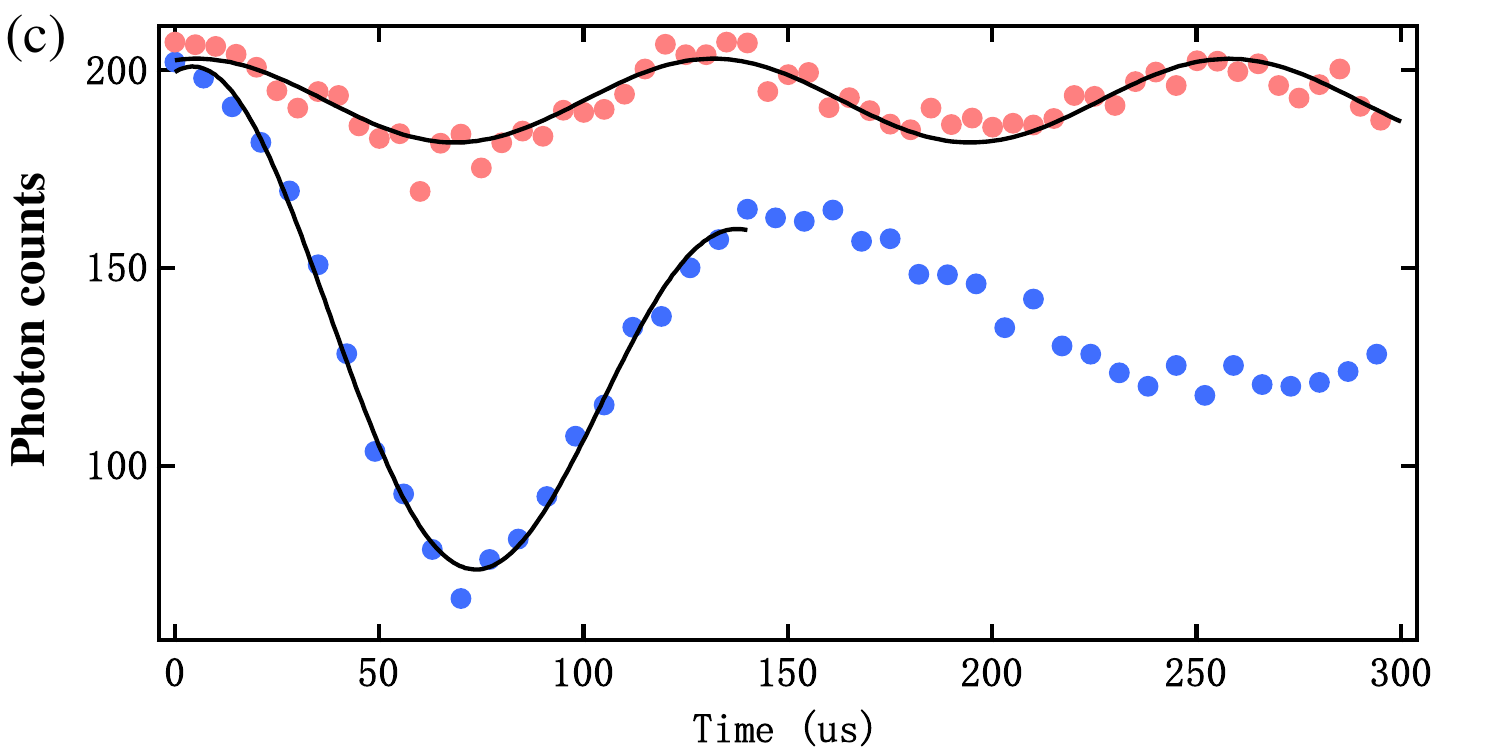} %
\end{minipage}
\caption{(a) Simulation of Rabi oscillations for blue (blue curve) and red-sideband
(red curve) transitions of the Radial COM mode with average phonon
number of 0.5. The black dots show the $\pi$ pulse positions of the
Rabi oscillations, which is used to calculated the sideband asymmetry
$p_{r}/(p_{b}-p_{r})$. (b) Simulated average phonon number as a function
of the sideband asymmetry. The blue squares are the simulation rusults
and the red line shows the linear fit for extracting the correction
factor. (c) Experimental results for Rabi oscillation of the radial asymmetry mode. The transition amplitude for blue-and red-sideband cooling is found to be 129 and 21 from the fitting and
the average phonon number is calculated to be 0.40 based the correction
factor found in Table. \ref{tab:tableS1}. }
\label{fig:figureS3}
\end{figure}

By numerically solving the Schr\"{o}dinger equations of red and blue sideband
transitions under the initial state $\ket{\downarrow\downarrow...\downarrow,n_{k}}$,
where $\ket{n_{k}}$ is the Fock state of mode $k$, we can obtain
the time evolution of the upstate of the ions $P_{\uparrow,n_{k}}(t)$
\cite{reve8,reve9}. Assuming the thermal state
distribution for the phonons, we can find the time evolution of $P_{\uparrow,n_{k}}(t)$
as a funtion of the average phonon number $\bar{n}_{k}$. Similar
to the single ion case, we can use the sideband spectroscopy and a
mode dependent factor to get the approximate evalution of average
phonon number $\bar{n}_{k}$. The factor can be extracted by fitting
the numerical found $p_{r}/(p_{b}-p_{r})$ versus $\bar{n}_{k}$.
Considering 4 ions with trap frequencies $\{\omega_{z},\omega_{x},\omega_{y}\}/2\pi=\{0.6,1.706,1.754\}$
MHz, the mode dependent factor is found as shown in table.\ref{tab:tableS1},
Fig. \ref{fig:figureS3}(b) is the simulated Rabi oscillation for
the red and blue sideband transition of radial COM mode with aveage
phonon number of 0.5. Fig. \ref{fig:figureS3}(b) shows the linear
fit for extracting the correction factor of the radial COM mode, and
the correction factor is found to be 2.10. As an example we extract the phonon number of the radial asymmetry mode by fitting the blue- and red-sideband Rabi oscillation as shown in Fig. \ref{fig:figureS3}(c). Based on the amplitude of the fitting function and correction factor above, the final phonon number is calculated to be 0.4.

\end{document}